\newcount\driver \newcount\mgnf \newcount\tipi
\newskip\ttglue
\def\TIPITOT{
\font\dodicirm=cmr12
\font\dodicii=cmmi12
\font\dodicisy=cmsy10 scaled\magstep1
\font\dodiciex=cmex10 scaled\magstep1
\font\dodiciit=cmti12
\font\dodicitt=cmtt12
\font\dodicibf=cmbx12 scaled\magstep1
\font\dodicisl=cmsl12
\font\ninerm=cmr9
\font\ninesy=cmsy9
\font\eightrm=cmr8
\font\eighti=cmmi8
\font\eightsy=cmsy8
\font\eightbf=cmbx8
\font\eighttt=cmtt8
\font\eightsl=cmsl8
\font\eightit=cmti8
\font\seirm=cmr6
\font\seibf=cmbx6
\font\seii=cmmi6
\font\seisy=cmsy6
\font\dodicitruecmr=cmr10 scaled\magstep1
\font\dodicitruecmsy=cmsy10 scaled\magstep1
\font\tentruecmr=cmr10
\font\tentruecmsy=cmsy10
\font\eighttruecmr=cmr8
\font\eighttruecmsy=cmsy8
\font\seventruecmr=cmr7
\font\seventruecmsy=cmsy7
\font\seitruecmr=cmr6
\font\seitruecmsy=cmsy6
\font\fivetruecmr=cmr5
\font\fivetruecmsy=cmsy5
\textfont\truecmr=\tentruecmr
\scriptfont\truecmr=\seventruecmr
\scriptscriptfont\truecmr=\fivetruecmr
\textfont\truecmsy=\tentruecmsy
\scriptfont\truecmsy=\seventruecmsy
\scriptscriptfont\truecmr=\fivetruecmr
\scriptscriptfont\truecmsy=\fivetruecmsy
\def \ottopunti{\def\rm{\fam0\eightrm}% switch to 8-point type
\textfont0=\eightrm \scriptfont0=\seirm \scriptscriptfont0=\fiverm
\textfont1=\eighti \scriptfont1=\seii   \scriptscriptfont1=\fivei
\textfont2=\eightsy \scriptfont2=\seisy   \scriptscriptfont2=\fivesy
\textfont3=\tenex \scriptfont3=\tenex   \scriptscriptfont3=\tenex
\textfont\itfam=\eightit  \def\it{\fam\itfam\eightit}%
\textfont\slfam=\eightsl  \def\sl{\fam\slfam\eightsl}%
\textfont\ttfam=\eighttt  \def\tt{\fam\ttfam\eighttt}%
\textfont\bffam=\eightbf  \scriptfont\bffam=\seibf
\scriptscriptfont\bffam=\fivebf  \def\bf{\fam\bffam\eightbf}%
\tt \ttglue=.5em plus.25em minus.15em
\setbox\strutbox=\hbox{\vrule height7pt depth2pt width0pt}%
\normalbaselineskip=9pt
\let\sc=\seirm  \let\big=\eightbig  \normalbaselines\rm
\textfont\truecmr=\eighttruecmr
\scriptfont\truecmr=\seitruecmr
\scriptscriptfont\truecmr=\fivetruecmr
\textfont\truecmsy=\eighttruecmsy
\scriptfont\truecmsy=\seitruecmsy
}\let\nota=\ottopunti}
\newfam\msbfam   %per uso in \TIPITOT
\newfam\truecmr  %per uso in \TIPITOT
\newfam\truecmsy %per uso in \TIPITOT
\def\TIPI{
\font\eightrm=cmr8
\font\eighti=cmmi8
\font\eightsy=cmsy8
\font\eightbf=cmbx8
\font\eighttt=cmtt8
\font\eightsl=cmsl8
\font\eightit=cmti8
\font\tentruecmr=cmr10
\font\tentruecmsy=cmsy10
\font\eighttruecmr=cmr8
\font\eighttruecmsy=cmsy8
\font\seitruecmr=cmr6
\textfont\truecmr=\tentruecmr
\textfont\truecmsy=\tentruecmsy
\def \ottopunti{\def\rm{\fam0\eightrm}% switch to 8-point type
\textfont0=\eightrm
\textfont1=\eighti
\textfont2=\eightsy
\textfont3=\tenex \scriptfont3=\tenex   \scriptscriptfont3=\tenex
\textfont\itfam=\eightit  \def\it{\fam\itfam\eightit}%
\textfont\slfam=\eightsl  \def\sl{\fam\slfam\eightsl}%
\textfont\ttfam=\eighttt  \def\tt{\fam\ttfam\eighttt}%
\textfont\bffam=\eightbf
\def\bf{\fam\bffam\eightbf}%
\tt \ttglue=.5em plus.25em minus.15em
\setbox\strutbox=\hbox{\vrule height7pt depth2pt width0pt}%
\normalbaselineskip=9pt
\let\sc=\seirm  \let\big=\eightbig  \normalbaselines\rm
\textfont\truecmr=\eighttruecmr
\scriptfont\truecmr=\seitruecmr
}\let\nota=\ottopunti}
\def\TIPIO{
\font\setterm=amr7 %\font\settei=ammi7
\font\settesy=amsy7 \font\settebf=ambx7 %\font\setteit=amit7
\def \settepunti{\def\rm{\fam0\setterm}% passaggio a tipi da 7-punti
\textfont0=\setterm   %\textfont1=\settei
\textfont2=\settesy   %\textfont3=\setteit
\textfont\bffam=\settebf  \def\bf{\fam\bffam\settebf}
\normalbaselineskip=9pt\normalbaselines\rm
}\let\nota=\settepunti}
\let\a=\alpha \let\b=\beta   \let\d=\delta  \let\e=\varepsilon
\let\f=\varphi      \let\k=\kappa  \let\l=\lambda
      \let\o=\omega      
\let\r=\rho  \let\s=\sigma

     \let\L=\Lambda 
      \let\X=\Xi

\def\data{\number\day/\ifcase\month\or gennaio \or febbraio \or marzo \or
aprile \or maggio \or giugno \or luglio \or agosto \or settembre
\or ottobre \or novembre \or dicembre \fi/\number\year;\,\the\time}

\def\data1{\number\day/\ifcase\month\or gennaio \or febbraio \or marzo \or
aprile \or maggio \or giugno \or luglio \or agosto \or settembre
\or ottobre \or novembre \or dicembre \fi/\number\year}

\newcount\pgn \pgn=1
\def\foglio{\number\numsec:\number\pgn
\global\advance\pgn by 1}
\def\foglioa{A\number\numsec:\number\pgn
\global\advance\pgn by 1}

\global\newcount\numsec\global\newcount\numfor
\global\newcount\numfig
\gdef\profonditastruttura{\dp\strutbox}

\def\senondefinito#1{\expandafter\ifx\csname#1\endcsname\relax}

\def\SIA #1,#2,#3 {\senondefinito{#1#2}%
\expandafter\xdef\csname #1#2\endcsname{#3}\else
\write16{???? ma #1,#2 e' gia' stato definito !!!!} \fi}

\def\etichetta(#1){(\veroparagrafo.\veraformula)%
\SIA e,#1,(\veroparagrafo.\veraformula) %
\global\advance\numfor by 1%
\write15{\string\FU (#1){\equ(#1)}}%
%\write16{ EQ \equ(#1) <==#1  }
}

\def\FU(#1)#2{\SIA fu,#1,#2 }

\def\etichettaa(#1){(A.\veraformula)%
\SIA e,#1,(A.\veraformula) %
\global\advance\numfor by 1%
\write15{\string\FU (#1){\equ(#1)}}%
%\write16{ EQ \equ(#1) <== #1  }
}

\def\getichetta(#1){Fig. \verafigura
\SIA e,#1,{\verafigura} %
\global\advance\numfig by 1%
\write15{\string\FU (#1){\equ(#1)}}%
\write16{ Fig. \equ(#1) ha simbolo  #1  }}

\newdimen\gwidth

\def\BOZZA{
\def\alato(##1){%
 {\vtop to \profonditastruttura{\baselineskip
 \profonditastruttura\vss
 \rlap{\kern-\hsize\kern-1.2truecm{$\scriptstyle##1$}}}}}
\def\galato(##1){\gwidth=\hsize \divide\gwidth by 2%
 {\vtop to \profonditastruttura{\baselineskip
 \profonditastruttura\vss
 \rlap{\kern-\gwidth\kern-1.2truecm{$\scriptstyle##1$}}}}}
}

\def\alato(#1){}
\def\galato(#1){}

\def\veroparagrafo{\number\numsec}\def\veraformula{\number\numfor}
\def\verafigura{\number\numfig}

\def\geq(#1){\getichetta(#1)\galato(#1)}
\def\Eq(#1){\eqno{\etichetta(#1)\alato(#1)}}
\def\eq(#1){\etichetta(#1)\alato(#1)}
\def\Eqa(#1){\eqno{\etichettaa(#1)\alato(#1)}}
\def\eqa(#1){\etichettaa(#1)\alato(#1)}
\def\eqv(#1){\senondefinito{fu#1}$\clubsuit$#1
\write16{#1 non e' (ancora) definito}%
\else\csname fu#1\endcsname\fi}
\def\equ(#1){\senondefinito{e#1}\eqv(#1)\else\csname e#1\endcsname\fi}

\def\include#1{
\openin13=#1.aux \ifeof13 \relax \else
\input #1.aux \closein13 \fi}

\openin14=\jobname.aux \ifeof14 \relax \else
\input \jobname.aux \closein14 \fi
\openout15=\jobname.aux
 
 \def\\{\noindent}

\def\tende#1{\vtop{\ialign{##\crcr\rightarrowfill\crcr
              \noalign{\kern-1pt\nointerlineskip}
              \hskip3.pt${\scriptstyle #1}$\hskip3.pt\crcr}}}
\def\otto{{\kern-1.truept\leftarrow\kern-5.truept\to\kern-1.truept}}

\def\mbox{\hbox}

\def\={{\equiv}}

\def\initfiat#1#2#3{
\mgnf=#1
\driver=#2
\tipi=#3
\ifnum\tipi=0\TIPIO \else\ifnum\tipi=1 \TIPI\else \TIPITOT\fi\fi
\ifnum\mgnf=0
\magnification=\magstep0\hoffset=0.cm
\voffset=-1truecm\hoffset=-.5truecm\hsize=16.5truecm \vsize=25.truecm
\baselineskip=14pt  % plus0.1pt minus0.1pt
\parindent=12pt
\lineskip=4pt\lineskiplimit=0.1pt      \parskip=0.1pt plus1pt
\def\ds{\displaystyle}\def\st{\scriptstyle}\def\sst{\scriptscriptstyle}
\font\seven=cmr7
\fi
\ifnum\mgnf=1
\magnification=\magstep1\hoffset=0.cm
\voffset=-1truecm\hoffset=-.5truecm\hsize=16.5truecm \vsize=24.truecm
\baselineskip=14pt  % plus0.1pt minus0.1pt
\parindent=12pt
\lineskip=4pt\lineskiplimit=0.1pt      \parskip=0.1pt plus1pt
\def\ds{\displaystyle}\def\st{\scriptstyle}\def\sst{\scriptscriptstyle}
\font\seven=cmr7
\fi
\setbox200\hbox{$\scriptscriptstyle \data1 $}
}
\initfiat {1}{1}{2}
\expandafter\ifx\csname amssym.def\endcsname\relax \else\endinput\fi
\expandafter\edef\csname amssym.def\endcsname{%
       \catcode`\noexpand\@=\the\catcode`\@\space}
\catcode`\@=11
\def\undefine#1{\let#1\undefined}
\def\newsymbol#1#2#3#4#5{\let\next@\relax
 \ifnum#2=\@ne\let\next@\msafam@\else
 \ifnum#2=\tw@\let\next@\msbfam@\fi\fi
 \mathchardef#1="#3\next@#4#5}
\def\mathhexbox@#1#2#3{\relax
 \ifmmode\mathpalette{}{\m@th\mathchar"#1#2#3}%
 \else\leavevmode\hbox{$\m@th\mathchar"#1#2#3$}\fi}
\def\hexnumber@#1{\ifcase#1 0\or 1\or 2\or 3\or 4\or 5\or 6\or 7\or 8\or
 9\or A\or B\or C\or D\or E\or F\fi}

\font\tenmsa=msam10
\font\sevenmsa=msam7
\font\fivemsa=msam5
\newfam\msafam
\textfont\msafam=\tenmsa
\scriptfont\msafam=\sevenmsa
\scriptscriptfont\msafam=\fivemsa
\edef\msafam@{\hexnumber@\msafam}
\mathchardef\dabar@"0\msafam@39
\def\dashrightarrow{\mathrel{\dabar@\dabar@\mathchar"0\msafam@4B}}
\def\dashleftarrow{\mathrel{\mathchar"0\msafam@4C\dabar@\dabar@}}

\def\ulcorner{\delimiter"4\msafam@70\msafam@70 }
\def\urcorner{\delimiter"5\msafam@71\msafam@71 }
\def\llcorner{\delimiter"4\msafam@78\msafam@78 }
\def\lrcorner{\delimiter"5\msafam@79\msafam@79 }
\def\yen{{\mathhexbox@\msafam@55 }}
\def\checkmark{{\mathhexbox@\msafam@58 }}
\def\circledR{{\mathhexbox@\msafam@72 }}
\def\maltese{{\mathhexbox@\msafam@7A }}

\font\tenmsb=msbm10
\font\sevenmsb=msbm7
\font\fivemsb=msbm5
\newfam\msbfam
\textfont\msbfam=\tenmsb
\scriptfont\msbfam=\sevenmsb
\scriptscriptfont\msbfam=\fivemsb
\edef\msbfam@{\hexnumber@\msbfam}
\def\Bbb#1{{\fam\msbfam\relax#1}}
\def\widehat#1{\setbox\z@\hbox{$\m@th#1$}%
 \ifdim\wd\z@>\tw@ em\mathaccent"0\msbfam@5B{#1}%
 \else\mathaccent"0362{#1}\fi}
\def\widetilde#1{\setbox\z@\hbox{$\m@th#1$}%
 \ifdim\wd\z@>\tw@ em\mathaccent"0\msbfam@5D{#1}%
 \else\mathaccent"0365{#1}\fi}
\font\teneufm=eufm10
\font\seveneufm=eufm7
\font\fiveeufm=eufm5
\newfam\eufmfam
\textfont\eufmfam=\teneufm
\scriptfont\eufmfam=\seveneufm
\scriptscriptfont\eufmfam=\fiveeufm

\csname amssym.def\endcsname
\def\sqr#1#2{{\vcenter{\vbox{\hrule height.#2pt
     \hbox{\vrule width.#2pt height#1pt \kern#1pt
   \vrule width.#2pt}\hrule height.#2pt}}}}

\def\11{\hbox{l}\!\!\!1\,}

\font\tenib=cmmib10
\newfam\mitbfam
\textfont\mitbfam=\tenib
\scriptfont\mitbfam=\seveni
\scriptscriptfont\mitbfam=\fivei

 %bold eta

\def\and{ \hbox{ and } }

\def\pa{\parallel}
\def\pt{\partial}
\def\l{\lambda}
\def\L{\Lambda}
\def\e{\varepsilon}
\def\a{\alpha}

\def\d{\delta}

\def\o{\omega}

\def\s{\sigma}

\def\newpage{\vfill\eject}
\def\Cal{\cal}
\def\to{\rightarrow}

\def\frac{\over}
\def\\{\cr}
\def\ref{}

\hbox{}
\vfill
\baselineskip12pt
\overfullrule=0in
\def\s{\sigma}

\def \ga {\gamma}  
\def \a {\alpha}

\def \de {\delta}  
 
\def \ti {\tilde}

\def \b {\beta}
\def \u {\underline}
\def \text {\hbox}
\newsymbol\gtrless 133F
\newsymbol\lessgtr 1337
\catcode`\X=12\catcode`\@=11
\def\n@wcount{\alloc@0\count\countdef\insc@unt}
\def\n@wwrite{\alloc@7\write\chardef\sixt@@n}
\def\n@wread{\alloc@6\read\chardef\sixt@@n}
\def\crossrefs#1{\ifx\alltgs#1\let\tr@ce=\alltgs\else\def\tr@ce{#1,}\fi
   \n@wwrite\cit@tionsout\openout\cit@tionsout=\jobname.cit 
   \write\cit@tionsout{\tr@ce}\expandafter\setfl@gs\tr@ce,}
\def\setfl@gs#1,{\def\@{#1}\ifx\@\empty\let\next=\relax
   \else\let\next=\setfl@gs\expandafter\xdef
   \csname#1tr@cetrue\endcsname{}\fi\next}
\newcount\sectno\sectno=0\newcount\subsectno\subsectno=0\def\r@s@t{\relax}
\def\resetall{\global\advance\sectno by 1\subsectno=0
  \gdef\firstpart{\number\sectno}\r@s@t}
\def\resetsub{\global\advance\subsectno by 1
   \gdef\firstpart{\number\sectno.\number\subsectno}\r@s@t}
\def\v@idline{\par}\def\firstpart{\number\sectno}
\def\l@c@l#1X{\firstpart.#1}\def\gl@b@l#1X{#1}\def\t@d@l#1X{{}}
\def\m@ketag#1#2{\expandafter\n@wcount\csname#2tagno\endcsname
     \csname#2tagno\endcsname=0\let\tail=\alltgs\xdef\alltgs{\tail#2,}%
  \ifx#1\l@c@l\let\tail=\r@s@t\xdef\r@s@t{\csname#2tagno\endcsname=0\tail}\fi
   \expandafter\gdef\csname#2cite\endcsname##1{\expandafter
 %the following line was replaced by the subseqent one, DNA 7/6/89
  %  \ifx\csname#2tag##1\endcsname\relax?\else\csname#2tag##1\endcsname\fi
     \ifx\csname#2tag##1\endcsname\relax?\else{\rm\csname#2tag##1\endcsname}\fi
    \expandafter\ifx\csname#2tr@cetrue\endcsname\relax\else
     \write\cit@tionsout{#2tag ##1 cited on page \folio.}\fi}%
   \expandafter\gdef\csname#2page\endcsname##1{\expandafter
     \ifx\csname#2page##1\endcsname\relax?\else\csname#2page##1\endcsname\fi
     \expandafter\ifx\csname#2tr@cetrue\endcsname\relax\else
     \write\cit@tionsout{#2tag ##1 cited on page \folio.}\fi}%
   \expandafter\gdef\csname#2tag\endcsname##1{\global\advance
     \csname#2tagno\endcsname by 1%
   \expandafter\ifx\csname#2check##1\endcsname\relax\else%
\fi%      \immediate\write16{Warning: #2tag ##1 used more than once.}\fi
   \expandafter\xdef\csname#2check##1\endcsname{}%
   \expandafter\xdef\csname#2tag##1\endcsname
     {#1\number\csname#2tagno\endcsnameX}%
   \write\t@gsout{#2tag ##1 assigned number \csname#2tag##1\endcsname\space
      on page \number\count0.}%
   \csname#2tag##1\endcsname}}%
\def\m@kecs #1tag #2 assigned number #3 on page #4.%
   {\expandafter\gdef\csname#1tag#2\endcsname{#3}
   \expandafter\gdef\csname#1page#2\endcsname{#4}}
\def\re@der{\ifeof\t@gsin\let\next=\relax\else
    \read\t@gsin to\t@gline\ifx\t@gline\v@idline\else
    \expandafter\m@kecs \t@gline\fi\let \next=\re@der\fi\next}
\def\t@gs#1{\def\alltgs{}\m@ketag#1e\m@ketag#1s\m@ketag\t@d@l p
    \m@ketag\gl@b@l r \n@wread\t@gsin\openin\t@gsin=\jobname.tgs \re@der
    \closein\t@gsin\n@wwrite\t@gsout\openout\t@gsout=\jobname.tgs }
\outer\def\localtags{\t@gs\l@c@l}
\outer\def\globaltags{\t@gs\gl@b@l}
\outer\def\newlocaltag#1{\m@ketag\l@c@l{#1}}
\outer\def\newglobaltag#1{\m@ketag\gl@b@l{#1}}

\def\t@gsoff#1,{\def\@{#1}\ifx\@\empty\let\next=\relax\else\let\next=\t@gsoff
   \expandafter\gdef\csname#1cite\endcsname{\relax}
   \expandafter\gdef\csname#1page\endcsname##1{?}
   \expandafter\gdef\csname#1tag\endcsname{\relax}\fi\next}
\def\verbatimtags{\let\ift@gs=\iffalse\ifx\alltgs\relax\else
   \expandafter\t@gsoff\alltgs,\fi}
\catcode`\X=11 \catcode`\@=\active
\localtags
\def \st{\scriptstyle}
\def \text{\hbox}
\def \\{\cr}
\def \de{\delta}
\def \u{\underline}
\def \Pa {\big |\big |}
\tolerance=10000
\magnification 1200
\baselineskip=12pt
\centerline {\dodicibf Solutions to the Boltzmann equation}
\centerline {\dodicibf in the Boussinesq regime.}
\vskip.7cm
\centerline{R. Esposito 
\footnote{$^1$} {\eightrm  Dipartimento di Matematica,  
Universit\'a degli Studi di L'Aquila, Coppito, 67100 L'Aquila, Italy},
\hskip .2cm
R. Marra 
\footnote{$^2$}{\eightrm Dipartimento di Fisica, Universit\`a di
Roma Tor Vergata, 00133 Roma, Italy.} 
\hskip.1cm and \hskip.1cm
J. L. Lebowitz  
\footnote{$^3$}{\eightrm Departments of Mathematics and
Physics,  Rutgers University, New Brunswick, NJ
08903, USA} 
}
\vskip .8cm
\noindent {\bf Abstract} We consider a gas in a horizontal slab, in which 
the top and bottom walls are kept at different temperatures. The system is described
by the Boltzmann equation (BE) with Maxwellian boundary conditions specifying the wall
temperatures. We study the behavior of the system when the Knudsen number $\e$ is
small and the temperature difference between the walls as well as the velocity field
is of order $\e$, while the gravitational force is of order $\e^2$.  We prove that
there exists a solution to the BE for $t\in (0,\bar t)$ which is near a global
Maxwellian, and whose moments are close, up to order $\e^2$ to the density,
velocity and temperature obtained from the smooth solution of the Oberbeck-Boussinesq
equations assumed to exist for $t\le \bar t$. 
\baselineskip=18pt
\vskip1cm
{\bf 1. Introduction.}
\vskip.3cm
\numsec= 1
\numfor= 1

In the study of thermal convection phenomena  the following system plays a 
paradigmatic role:  a  viscous heat conducting fluid between flat horizontal plates 
with the lower plate maintained at a temperature greater than the upper one. When the
temperature difference between the plates is small, the stationary state  is one in
which the fluid is at rest with a linear temperature profile. When the temperature
difference is made larger, the gravitational buoyancy force acting on the light,
higher  temperature fluid below, overcomes the effects of viscosity and a new
stationary state of thermal Rayleigh-Benard convection sets in. In typical
experimental situations the variations of temperature and density are small and 
the system is described in the Boussinesq approximation under which the Navier-Stokes 
equations reduce to the Oberbeck-Boussinesq equations (OBE). This approximation,
which is formulated on a phenomenological basis, gives quantitatively correct
predictions in most cases [\rcite{DR}].

A justification  of this approximation, based on introducing a 
scaling which leaves the Rayleigh number invariant is given in [\rcite{EM}] (see also
[\rcite{Mi}] and [\rcite{Jo}]). This approach follows the general strategy of taking
into account the invariance, under appropriate scaling, of the hydrodynamical 
equations. Such considerations allow, for example, to derive the incompressible
Navier-Stokes equations from the compressible ones, as well as from microscopic and
kinetic models.

The main aim of this paper is to derive the OBE starting from the Boltzmann equation
(BE), which describes gases on the kinetic level, intermediate
between the microscopic and the macroscopic. To go from the kinetic BE to a
hydrodynamical one it is necessary to consider situations in which the Knudsen number
$\e$, the ratio between the mean free path and the size of the slab, is very small. 
It is well known that the inviscid Euler equations correctly describe the behavior 
of this system for times of order $\e^{-1}$ in the limit $\e\to 0$ [\rcite{Ca}].  To
obtain the OBE we need to consider longer times, of order $\e^{-2}$, so as to get the
effects of viscosity  and thermal conductivity. To make this possible, we have  to
study the system in the incompressible regime, corresponding to macroscopic velocity
fields of order $\e$ [\rcite{DEL}]. This scaling would appear, at first sight, to
require that the force $G$ be scaled as $\e^3$ to be consistent with the
incompressible regime, i.e. to get a finite force term in the 
equation for the velocity field. However, the case of a conservative force
is special from this point of view in that larger forces are permissible. In fact
we find that $G$ has to be scaled as $\e^2$ in order to keep  the Rayleigh number
finite. 

We take the walls to be at fixed temperatures and impose a no-slip boundary condition
for the velocity field at the hydrodynamic level. This is modeled at the kinetic level
by assuming that each particle colliding with a wall is reflected with a  random
velocity, distributed according to the equilibrium distribution at the temperature
of that wall, i.e. we assume Maxwellian boundary conditions.

Our solution of the BE is given in terms of a truncated expansion in $\e$ whose 
leading term in the bulk is a global Maxwellian. The term of order $\e$, denoted
by $f_1$, determines the hydrodynamic quantities which are close, up to order $\e^2$,
to the density, velocity and temperature solutions of the OBE. Near the boundary, in
a thin layer of size $\e$, the hydrodynamical approximation is not correct and we
provide a detailed description of the solution in this region. The initial datum is
chosen to match the expansion up to order $\e^2$,  to avoid a treatment of the
initial layer which is more or less standard.

The validity of the expansion is established up to a time $\bar t$ such that the OBE
have a sufficiently regular  solution by estimating the remainder. This expansion
technique goes back to Hilbert, Chapmann and Enskog;  a rigorous proof of the
hydrodynamic limit is given for the Euler case in [\rcite{Ca}] and for the
incompressible Navier-Stokes case in [\rcite{DEL}].  In particular, in the absence of
gravity, the results of the present paper extend the ones of [\rcite{DEL}] to the
case of a fluid confined in a domain with walls at different
temperatures  modeled by Maxwellian boundary conditions.
The main technical difficulty for such systems, even in the absence of an
external force, is in dealing with the  terms  coming from the boundary
conditions in the estimate for the remainder. The approach proposed in [\rcite{Ca}],
and used also in [\rcite{DEL}], is based on the estimate of some Sobolev norm of
the solutions. But in the presence of the boundaries this cannot be used because  the
derivatives of the solution may become singular at the boundaries. We avoid the
estimates of the derivatives by first looking for $L_2$ estimates and then improving
them to $L_\infty$ estimates. This technique was already used in [\rcite{ELM1}] and
[\rcite{ELM2}] which concern essentially one-dimensional problems, i.e.\ a
compressible fluid in a slab with the walls held at fixed temperatures
under the action of a force parallel to the walls in the stationary regime. Since we consider
here the fully three dimensional time dependent case we need to modify the method to
bound the $L_\infty$ norm of the remainder by using a new method based on a result in
[\rcite{UA}]. This is presented in Section 4 and in the Appendix. A formal expansion including boundary layer corrections was given earlier in
[\rcite{BCN1}]. 

The case with a force perpendicular to the walls  presents  extra difficulties
stemming from the fact that we need good properties of the derivatives with respect
to $v_z$, the vertical component of the velocity, to get the exponential  decay of
certain boundary layer terms and to control the remainder. Unfortunately, the
derivative with respect $v_z$ is  singular at the boundaries at $v_z=0$. To overcome
this difficulty we have to decompose the force into a part acting in the bulk only
and a part acting only near the boundaries, decreasing to zero at a distance of order
$\e$. The Milne problem we consider for the boundary layer terms  involve these short
range  forces and can be solved by using the result in [\rcite{CEM}], where it is
proven that the $v_z$ derivative is bounded in $L_2$ and $L_\infty$ norms locally,
away from the boundaries. This is enough to control the terms appearing in the
equation for the remainder and in the Milne problems for the higher order corrections.

While the above results are independent of the nature of the solution of the OBE, we
are only able to construct stationary solutions to the Boltzmann equation which
correspond, at the hydrodynamic level, to the behavior of the purely conducting
stationary solution of OBE. This is due to the fact that
the methods in this paper are based on the perturbation of a
global Maxwellian and therefore require that the temperature
difference between the plates be small, corresponding to having
a small Rayleigh number. We expect to be able to construct the
purely conductive solution of the Boltzmann equation for any
Rayleigh number, even when the basic hydrodynamic solution
becomes unstable,  by perturbing a Maxwellian corresponding to 
the hydrodynamic solution. This involves many technical
difficulties and will be discussed in a forthcoming paper. The
hope is to extend these results to the convective solutions
which appear for larger values of the Railegh number.

\vskip.5cm

{\bf 2. Hydrodynamic description.}
 
\numsec=2
\numfor=1
\vskip .5cm

We consider an incompressible heat-conducting viscous fluid in a horizontal slab
$\Lambda=\Bbb T_L^2\times [-1,1]$, where $\Bbb T_L^2$ is the two dimensional torus  
of size $L$. The acceleration of gravity is the vector $\u G=(0,0,-G)$. We specify 
the temperature on the boundaries $z=\pm 1$ as: 
$$T(1)=T_+, \quad \quad  T(-1)=T_-=T_++\delta T.
\quad\quad\Eq(2.0)$$  The total mass of the fluid is 
specified to be $m$ with   $\bar \rho= m|\Lambda|^{-1}$ the
corresponding mass density. We will assume
$T_-\ge T_+$. Denote 
$\theta=T-T_-$  the deviation of the temperature from $T_-$ and set
$$\tilde\theta= \theta-{2\over 5}G(1+z).$$

The equations describing the evolution of the velocity and
temperature field, $u$ and
$\tilde\theta$, are the Oberbeck-Boussinesq (OBE) equations (see [\rcite{Bo}],
[\rcite{Jo}]), which we write as
$$\eqalign{&\text{div} u=0,\\
&\bar\rho (\partial_t u+u\cdot \nabla u)=\eta \Delta u
-\nabla \tilde p -\alpha \bar\rho\tilde\theta \u G\\
&{5\over 2}\bar\rho(\partial_t \tilde\theta+u\cdot \nabla
\tilde\theta)= \k \Delta
\tilde\theta.}\Eq(2.12)$$ 
Here $\eta$ is the kinematic viscosity coefficient, $\k$ the heat conduction
coefficient, $\a=T_-^{-1}$ the coefficient of thermal expansion,
$$\tilde p=  p -{3\over 10}{\bar\rho\over R T_-}G^2(1+z)^2$$
and $ p$ is the unknown pressure which arises from the incompressibility
constraint.
The initial conditions are
$$u(\u x, 0)=u_0(\u x),\quad\quad \tilde\theta(\u x, 0)=\theta_0(\u
x)-{2\over 5}G(1+z)\Eq(ic)$$ 
for any $\u x\in \L$, with $\text{div} u_0=0$.
The boundary conditions for this problem are
$$u(x,y,-1,t)=u(x,y,1,t)=0,\quad\quad \tilde\theta(x,y,-1,t)=0, \quad
\tilde\theta(x,y,1,t)=-\d T-{4\over 5} G,\Eq(bc)$$
for any $(x,y)\in\Bbb T_L^2$ and any positive $t$.

Moreover, let $r=\r-\bar \r$ be the deviation of the density from the homogeneous
density $\bar \r$, let $\tilde r=r+\bar\r G(1+z)/T_-$. Then $\tilde r$ is determined
by the Boussinesq condition 
$$\bar\rho \nabla  \theta + T_- \nabla \tilde r=0\Eq(2.10)$$
\vskip.2cm
It can be proved that if $u_0$ and $\theta_0$ are smooth functions of $\u x$
(i.e. with Sobolev $H_s(\L)$-norm finite for some $s$ sufficiently large), then there
is
$\bar t$ depending on the initial and boundary data such that the system \equ(2.12),
\equ(ic),
\equ(bc) has an unique solution, at least as smooth as the initial data, for $0<t\le
\bar t$. We do not give the proof of this statement which is rather standard and
refer to  [\rcite{Jo}] for details.

The regime under which the OBE are expected to be
valid 
correspond to a low Mach
number, a 
sufficiently weak gravity  and a small difference between the temperatures
of the top and bottom walls.  To make this percise we introduce a space scale parameter
$\e$ and rescale the variables as follows:
$$\u x\to \e^{-1}\u x,\quad t\to \e^{-2}t,\quad u\to\e u,\quad \theta\to\e\theta,
\quad G\to\e^2 G,\quad\delta T\to\e\delta T \Eq(2.1)$$ 
This scaling is a natural one in order to derive the OBE because, under it, the
Rayleigh number $Ra$ defined as 
$$Ra = \left({GL^3\delta T\over \k\nu T_+}\right)^{1/2}.$$
is kept fixed.  
We refer to [\rcite{EM}] for the detailed (formal) derivation of \equ(2.12), \equ(ic),
\equ(bc), \equ(2.10) from the compressible Navier-Stokes in the limit
$\epsilon \to 0$, while in next sections we
will provide its rigorous derivation from the Boltzmann equation.

We note that our equations do  not coincide exactly with  the
usual Oberbeck-Boussinesq (OBE) equations as given in 
[\rcite{Bo}], [\rcite{Jo}] because of the term proportional to
$G$ in the boundary conditions for $\tilde \theta$ and of the
quadratic term in $G$ in the definition of $\tilde p$. In the
usual experimental conditions (see [\rcite{DR}]) such terms are 
much smaller than the others, so one can neglect the effect of
the variation of the density due to the gravitational force. If
we denote by $\rho_s$ and $T_s$  the solution of the stationary
problem
$${d\over dz}P_s=- G\rho_s,\quad\quad \Delta T_s=0,$$
with $P_s= \rho_s T_s$ and  boundary conditions \equ(2.0) the approximation
corresponds to setting $P_s\sim P(1)\equiv\r(1)T(1)$.   In this
way we would recover the usual OBE. 
 We finally remark that
the Boussinesq condition \equ(2.10),  which is  assumed as a ``equation of state''
in the usual discussions of the Boussinesq approximation (see [\rcite{Bo}],
[\rcite{Mi}]), in our approach is just a consequence of the scaling
limit (2.6). 
\goodbreak
\vskip1cm
\numsec=3
\numfor=1
\centerline {\bf 3. Kinetic description }
\vskip .5cm
We consider the BE for a gas between parallel planes. We keep the notations of 
Sect.~2. To model the  hydrodynamic boundary conditions we choose the so called
Maxwellian boundary conditions:  when a particle hits the walls of the slab ($z=-1$
or $z=1$) it is diffusely reflected  with a velocity distributed according to a
Maxwellian with zero mean velocity and prescribed temperatures $T_-$ and $T_+$
respectively. In the language of kinetic theory this means that
the {\it accommodation coefficient} equals one. The above
prescription implies  the impermeability  of the walls, namely
no particle flux  across the boundary is allowed. We introduce
as  scale parameter $\e$ the Knudsen number. The height of the
slab is $2\e^{-1}$, hence in rescaled variables $z \in [-1,1]$. 
We take for simplicity periodic conditions in the $x,y$
direction, and call
$$\Lambda= \{\u x:(x,y)\in \Bbb T_L^2, z\in (-1,1)\}, \qquad  \Omega =\{(\u {x},v)\,|
\u x\in\Lambda, v\in \Bbb R ^3\}\Eq(3.0).$$ 

The BE rescaled according to \equ(2.1) is 
$$\partial_t f^\e +{\e}^{-1}v\cdot \nabla f^\e + \u G\cdot\nabla_v f^\e= {\e}^{-2}
Q(f^\e,f^\e),\Eq(3.1)$$
with
$$Q(f,f)(\u x,v,t)= \int_{\Bbb R^3}d\/v_*\int_{S_2}d\/\o
B(\o,|v-v_*|)\big\{f(\u x,v',t)f(\u x,v'_*,t)-f(\u x,v,t)f(\u x,v_*,t)\big\}$$
where $S_2=\{\o\in\Bbb R^3\,|\o^2=1\}$, $B$ is the differential cross section and
$v'$,$v'_*$ are the incoming velocities of a collision with outgoing velocities $v$,
$v_*$ and impact parameter $\o$. We confine ourselves to the collision cross section
$B(\o,V)=|V\cdot \o|$ corresponding to hard spheres
[\rcite{CIP}].  

The initial condition is
$$f^\e(\underline{x},v; 0)=f^\e_0(\underline{x},v), \phantom{....}
\u x\in \L,\Eq(3.2).$$ 
The precise form of $f^\e_0$ will be specified below where it
will be seen that it cannot be given  arbitrarily if one wants 
to avoid a detailed analysis of the initial layer. However, we
assume the initial datum  $f_0$ non negative and  normalized to
the total mass which we set to $1$. 

The boundary conditions are:
$$\eqalign{f^\e(x,y,-1,v; t)&=\alpha_- \overline M_-(v), \phantom{....}
v_z>0, \phantom{..} t>0,\\
f^e(x,y,1,v; t)&=\alpha_+ \overline M_+(v), \phantom{....}
 v_z<0,\phantom{..}t>0,}
\Eq(3.4) $$
with  
$${\overline M_\pm}(v) = {1 \over 2\pi
T_\pm^2} \text{\/e}^{-v^2/2T_\pm}, \Eq(3.5)$$
normalized so that $\int _{v_y \lessgtr 0} |v_y|{\overline M_\pm}(v) dv =1$. 
The temperature $T_+$ is assumed to satisfy
$$T_+=T_-(1-2\e \l)$$
with $\l$ independent of $\e$, according to the scaling \equ(2.1).

The quantities $\alpha_\pm$ must be chosen in such a way that the impermeability 
condition of the walls is assured, i.e.
$$ \langle v_z f^\e \rangle\equiv
\int_{\Bbb R^3} v_z f^\e dv= 0  \phantom{....}  \text{ for } z=\pm 1,
\Eq(3.6) $$
where we have introduced the notation $\langle f\rangle =\int_{\Bbb R^3}f(v)dv$.
Condition \equ(3.6) and the normalization of ${\overline M_\pm}$ 
imply: 
$$\a_\pm=\pm\int _{v_z \gtrless 0} v_z f^\e(x,y,\pm 1,v; t)dv \Eq(3.7)$$
Namely, $\alpha_{\pm}$ represent the outgoing (from the fluid) fluxes of mass in the
direction $z$. The impermeability condition implies that the normalization of the
solution to \equ(3.1) stays constant and therefore we will look for solutions to
\equ(3.1) which are normalized to $1$ as the initial datum. 

The macroscopic behavior should be  recovered in the limit $\e$ going to zero. More
precisely  we expect that for $\e$ small the behavior of the solution \equ(3.1) is
very close to the hydrodynamical one, in the sense that it can be described by a
local Maxwellian with parameters which differ from constants by terms  of order
$\e$, and that these terms are solution of the OBE. At higher order in $\e$ there will
both  be kinetic and boundary layer corrections. Therefore we look for a solution of
the form
$$f^\e=M + \e f_1 +\sum_{n=2}^7\e^n f_n +\e^4 R \Eq(3.8)$$
where 
$M$ is the global Maxwellian 
$$M( \bar\rho, 0, T_-;v)={\bar\rho\over(2\pi T_-)^{3/2}}
\text{e}^{-v^2/2T_-}.$$ If we put \equ(3.8) in the BE \equ(3.1) we see
immediately that  $f_1$ has to satisfy
$${\cal L} f_1:=2Q(M,f_1)=0,\Eq(3.9)$$
where ${\cal L}$ is the linearized Boltzmann operator.
\equ(3.9)  implies that $f_1$ has to be in $\text{Null}\  {\cal L}$, which means
that it is a combination of the
collision invariants $M\chi_i$ with $\chi_i(v)=1,v_i,(v^2-3
T_-)/2$, for
$i=0$, $i=1,2,3$ and $i=4$ respectively, suitably normalized  to form
an orthonormal set, in $L_2(M(v)^{-1}d\/v)$. Hence we have
$$f_1= M\sum_{i=0}^4 \chi_i t_i(t,\underline{x})\equiv
M\Big({ r\over \bar\rho} + {u\cdot v\over T_-} + 
\theta{|v|^2- 3 T_-\over 2 T_-^2}\Big).\Eq(3.10)$$
The  functions
$t_i(t,\underline{x})$ and/or $ r, u, \theta$ will satisfy
equations to be determined. To write the conditions for 
$f_n$ we decompose them into two parts  $B_n$ and $b^{\pm}_n$, representing the bulk
 and boundary layer corrections. The latter are significantly 
different from $0$ only near the boundary. The 
$B_n$ have to satisfy  for $n=2,\cdots,7$
$$\partial_t B_{n-2} +v\cdot\nabla B_{n-1} +\u G\cdot\nabla_v B_{n-2}=2Q(M,B_{n})
+\sum_{i+j=n}Q(B_i,B_j)\Eq(3.12)$$ 
where $B_0\equiv M$ and $B_1\equiv f_1$.

We note that the condition $f_1=B_1$ means that there is no boundary layer correction
to the first order in $\e$. To make this  compatible with \equ(3.4) we need to
assume that $u(x,y,-1,t)=u(x,y,1,t)=0$, $\theta(x,y,-1,t)=0$, $\theta(x,y,1,t)=2
\l T_-$ for any $(x,y)\in \Bbb T_L$ and any $t>0$. We remark
that $M+\e f_1$, when evaluated for $z=1$ is not proportional to
the Maxwellian $\bar M_+$, even with previous assumptions, but
differs from it for terms of order $\e^2$ which will appear
in the corrections of higer order. 

To construct the boundary layer terms we decompose the constant gravity force
$\underline{G}=(0,0,-G)$ into three parts: a bulk part
$\u G_0$  and two boundary parts $\u G^{\pm}$ which are different from zero in the
bulk and near the walls respectively. Their definition is
$$\u G=\u G^+ +\u G^0 +\u G^-$$
with  $\u G_0$ and $\u G^{\pm}$ smooth functions such that
$$
\u G^+(z)=\cases{& $\u G$, \quad $1-\delta\  \e\le z\le 1$\cr
& $0$, \quad $-1\le z\le 1-2\delta\ \e$},
\qquad
\u G=\cases{& $\u G$, \quad $-1\le z\le -1+\delta\ \e$\cr
& $0$, \quad $-1+2\delta\ \e\le z\le 1$}
\Eq(3.12.1)$$
$$
\hskip.4cm \u G(z)= \cases{& $\u G$, \quad $-1+2\delta\ \e\le z\le 1-2\delta\ \e$\cr
& $0$, \quad $|z|\ge 1-\delta\ \e$}.
\Eq(3.12.2)
$$
Moreover we have to scale back to microscopic coordinates around $z=\pm1$. Setting
$z^{\pm}=\e^{-1}(1\mp z)$ so that $z^{\pm}\in [0,2\e^{-1}]$ we have that
$\u  G^\pm(z^\pm)$ is zero for $z^\pm\in[2\delta,2\e^{-1}]$. The boundary layer
corrections relative to the wall $z=\pm 1$,  $b^{\pm}_n$, are chosen to satisfy, for
$n=2\dots 5$, the equations
$$\eqalign{&
v_z{\pt b^{\pm}_n\over \pt z^{\pm}}\mp\e^2 G^\pm{\pt\over\pt_{v_z} }b^{\pm}_{n} =
\partial_t  b^{\pm}_{n-2} +\hat v\cdot\hat\nabla b^{\pm}_{n-1} +
{\cal L}^\pm b^{\pm}_n +2Q(\Delta M,b^-_{n-1})\chi^\pm\\&
\mp(G^0+G^\mp) {\pt\over\pt{v_z} }b^{\pm}_{n-2}+\sum_{\st i,j\ge
1 \atop\st i+j=n }
\Big[2Q(B_i,b^{\pm} _j)
 + Q(b^{\pm} _i,b^{\pm} _j) +Q(b^{\mp} _i,b^{\mp} _j)\Big],
}\Eq(3.13)$$
where we have put
$$b^\pm_0=b^\pm_1=0,\qquad \hat v =(v_x, v_y),\qquad \hat \nabla=
(\pt_x, \pt_y),\qquad \chi^+=1, \chi^-=0$$
$${\cal L}^\pm=2Q(M^\pm,\cdot), \qquad\Delta 
M=\e^{-1}[M-M_+], \qquad M_+=M(\rho_+,0,T_+;v),$$
and $\r_+=\bar\r+\e r(1)$.

Finally the equation for the remainder is
$$\partial_t R+ {1\over \e}v\cdot\nabla R +\u G\cdot\nabla_v
R= {1\over \e^2}{{\cal L}} R + {1\over \e}{{\cal L}}^{(1)} R + 
{\cal L}^{(2)} R +
\e^2 Q(R,R) + \e^2 A \Eq(3.14)$$
with 
$${\cal L}^{(1)} R=2Q(f_1,R),\phantom{...}
{\cal L}^{(2)} R = 2Q(\sum_{n=2}^7 \e^{n-2}f_n,R)
\Eq(3.15)$$
and $A$ given by
$$\eqalign{ A =&- \pt_t(f_6 +\e f_7)- v\cdot\nabla B_7-\hat
v\cdot\hat\nabla b^+ _7-\hat v\cdot\hat\nabla b^- _7 - \u G\cdot\nabla_v (B_6 +\e
B_7)\\ & -  (\u G^0+\u G^-)\cdot\nabla_{v} [(b^+_6 +\e
b^+_7)]-(\u G^0+\u G^+)\cdot \nabla_{v}[(b^-_6 +\e b^-_7)]+\\ &
2Q(\Delta M, b^-) +\sum_{\st k,m\ge 1 \atop\st k+m\ge 8 }
\e^{k+m-8} Q(f_k,f_m)
}
\Eq(3.16)$$

The boundary conditions for these equations have to be chosen
in such a way as to satisfy  \equ(3.4)--\equ(3.6) for $f^\e$.
Since we are interested in the case $T_+ =T_-(1 - 2 \e \lambda)$ it is easy  to
satisfy \equ(3.4) up to the first order in $\e$, because M is already
a Maxwellian whose temperature and velocity field are chosen to fit
with $\overline M_-$, while $M+\e f_1$ is close to be
proportional to $\overline M_+$ at $z=1$, up to terms of
order $\e^2$. 

{}From the second order on we have to use 
boundary layer terms to fit boundary conditions. In fact,
as we will see later, the $B_n$, for $n\ge 2$, do not reduce
to $\alpha^\pm_n\overline M_\pm$. The idea is to introduce at one of
the boundaries, say $z=1$, 
the correction $b^+_2$  so that $B_2+b^+_2$ is proportional to 
$\overline M_+$ for $v_z<0$. On the other hand,  
the same has to be done at $z=-1$ and  $f_2$ is modified by $b^-_2$
also. This changes again $f_2$ at $z=1$ by non Maxwellian terms.
However, since $b^-_2$ decays exponentially fast, the
modification  is exponentially
small in $\e^{-1}$. Therefore we impose on
the $f_n$ the following boundary conditions: %
$$\eqalign{ f_n(\u x,v; t)&= \a^-_n  
\overline M_-(v) +\ga^-_{n,\e} (v), \phantom{..}z=-1, \phantom{....}
v_z>0, \phantom{..}t> 0\\ f_n(\u x,v; t)&=\a^+_n 
\overline M_+(v) + 
\ga^+ _{n,\e} (v), \phantom{..}z=1, \phantom{....}\  v_z<0, \phantom{..}t>
0}
\Eq(3.17) 
$$
with the functions $\ga^\pm _{n,\e} (v)$ exponentially small in
$\e^{-1}$ and such that $\langle\ga^\pm _{n,\e} v_z \rangle=0$, to be
specified later. Moreover
$$\alpha^\pm_n=\pm\int _{v_z \gtrless 0} v_z f_n(x,y,\pm 1,v; t)dv
\Eq(3.18)$$ 
Finally, to fulfil \equ(3.4) we impose the
following conditions on $R$:
$$ R(\u x,v; t)= \a^-_R \overline M_-(v)
-\sum_{n=2}^7  \e ^{n-3} \ga^- _{n,\e}, \phantom{..}z=-1, 
\phantom{....}
v_z>0,\phantom{..}t> 0\Eq(3.19)
$$
$$
 R(\u x,v; t)=\a^+_R \overline
M_+(v)  -\sum_{n=2}^7  \e ^{n-3} \ga^+_ {n,\e},\phantom{..}z=1, \phantom{....}
v_z<0,\phantom{..}t> 0
\Eq(3.20) $$

The initial conditions for $R({\u x},v; 0)$ are chosen to
be $R({\u x},v; 0)=0,\phantom{..} z\ne
\pm 1$ for simplicity. The initial values for the $f_n$'s are partly 
determined by the procedure below, so that only their
hydrodynamical part can be assigned arbitrarily. To remove such restrictions
one would have  to include an  analysis of the initial layer, which we skip to
make the presentation simpler. Finally we impose  the conditions
$$\int_{\Omega} d\u x dv f_n =0=\int_{\Omega} d\u x dv R \Eq(3.20.1)$$
to ensure the normalization of the solution. Note that this condition on
$R$ is satisfied because it is true at time $t=0$.

\vskip.3cm
\noindent$\underline{ \hbox{\rm Outline of Solution.}}$
\vskip.3cm
The equations for the $f_n$ are coupled in a complicated way and
have to be solved in the proper sequence, which we now outline.
The hydrodynamical part of the bulk terms is determined by the
solvability conditions for \equ(3.12), that we  get by multiplying \equ(3.12)
by $\chi_i, i=0\dots4$, integrating  over velocity and using the fact
that $\langle Q(f,g)\chi_i\rangle=0$. The solvability condition for
\equ(3.12) with $n=2$ is
$$
\langle \chi_i[v\cdot \nabla f_1 +\u G\cdot\nabla_v M]
\rangle=0, \quad i=0,\cdots,4
\Eq(3.22)$$ 
because the Maxwellian $M$ does not depend on $x$ and $t$. This is
equivalent to 
$$ 
{\text {div}} u=0, \phantom{...}\quad\quad
\bar\rho \nabla  \theta+ T_-\nabla  r=\bar\rho\/G.
 \Eq(3.23)
$$
The first one is the usual incompressibility  condition while the
second one becomes the Boussinesq condition \equ(2.10), when one
defines $\tilde r=r+\bar\rho({G/T_-} )(1+z)$.  Once
\equ(3.23) are satisfied, we can deduce from
\equ(3.12) with $n=2$ the following expression for $B_2$, where ${\cal L}^{-1}$ 
denotes the inverse of the restriction of ${\cal L}$ to the orthogonal of its null
space
$$
B_2={\cal L} ^{-1}\Big[v\cdot \nabla f_1 +\u G\cdot\nabla_v
M-Q(f_1,f_1)\Big] +M\sum_{i=0}^ 4\chi_i\  t^{(2)}_i(t,{\u x})
\Eq(3.24)
$$

The solvability condition for \equ(3.12) with $n=3$ is 
$$\langle\chi_i[\pt_t f_1  +
\u G\cdot\nabla_v f_1+ v\cdot \nabla B_2]\rangle=0,\quad
i=0,\cdots,4 \Eq(3.25)$$ and this produces the equations for $u$
and $\theta$.  Let us fix  $i=1, 2, 3$ in \equ(3.25). Then 
the first term gives the time derivative of $\bar\rho u$. The second  one
reduces to  $-\u G\/\hat r$ after integrating by parts. Finally we write 
$$ \langle v\otimes v B_2\rangle= \langle[v\otimes v - {v^2 \over 3}
\Bbb I] B_2\rangle +\langle{v^2 \over 3} \Bbb I B_2\rangle$$
The first term, as is well known,  
gives rise to the dissipative and transport terms 
in the second of \equ(2.12), while the second one is the
second order correction to the pressure $P_2$. The result is
$$\bar\rho(\pt_t u +u\cdot \nabla u)=
\nu \Delta u-\nabla P_2 + G r.$$
Using the Boussinesq condition, the definitions of $r$  and $\theta$
and the relation between $P_2$ and $p$ of Section 2, we find
$\text{\equ(2.12)}_2$  as in Section 2, with $\eta$ given by
$$\eta=\langle(v\otimes v - {v^2 \over 3}
\Bbb I){\cal L}^{-1}[M(v\otimes v - {v^2 \over 3}
\Bbb I)]\rangle.$$ 

To get the equation for the temperature, it is convenient  to
replace $\chi_4$ in \equ(3.25)  by $\hat\chi_4= {1\over 2}[v^2 -5T_-)]$. A simple
computation, using the Boussinesq condition,  yields:
$$\langle {1\over 2}[v^2 -5T_-)]f_1\rangle=
{5\over 2}\bar\rho \big[\theta - ({2\over 5} G )\big] +
\text{const.},$$
$$\langle {1\over 2}[v^2 -5T_-)\underline{G}\cdot\nabla_v f_1\rangle=
-\bar\rho u_z G,$$
$$\langle v{1\over 2}[v^2 -5T_-)]B_2\rangle=
-\k \nabla \theta + {5\over 2}\bar\rho u\/\theta(1-z)).$$
Collecting the above results  we get \equ(2.12) with $\k$
given by
$$\k= \langle v {1\over 2}(v^2 -5T_-)
{\cal L}^{-1}[M v{1\over 2}(v^2 -5T_-)]\rangle.\Eq(3.26)$$

Finally, equation \equ(3.25) with $i=0$ gives
$$\pt_t r={\text {div}}\  \underline{t}^{(2)}, \quad \quad 
\underline{t}^{(2)}=(t_1^{(2)},t_2^{(2)}, t_3^{(2)}).\Eq(3.27)$$

Summarizing our results so far: we have shown that, assuming $u, p, 
\theta$ satisfy the  OBE \equ(2.12), \equ(ic), \equ(bc),
\equ(2.10) up to a time $\bar t$, the coefficients $t_i$
entering in the definition of $f_1$ are determined.  Therefore,
once initial and boundary conditions for the OBE are specified,
$f_1$ is completely determined as a 
function of $(t,{\u x}, v)$.

On the other hand the hydrodynamic part of $B_2$ is not yet
determined, but, by \equ(3.27), ${\text {div}} \ \underline{t}^{(2)}$ is
determined in terms of $r$.  Moreover, a combination  of 
$t^{(2)}_0$ and $t^{(2)}_4$ contributes to the pressure $p$ which is
determined by the OBE, so that these parameters are not
independent. 

The non-hydrodynamic part of $B_2$ depends  on  
the derivatives of $r, u, \theta$ which are in general
different from zero on the boundaries. Therefore $B_2$ violates the
boundary conditions and we need to introduce $b^{\pm}_2$ to
adjust the boundary conditions. We choose $b^-_2$
by solving, for any $t>0$, the Milne problem
$$v_z{\pt\over\pt z} h -\e^2 G^-{\pt\over\pt v_z}h={\cal L}^- h, \phantom{...}\langle
v_z  h 
\rangle=0$$
with boundary condition (at $z^-=0$)  prescribing 
the incoming flux as the opposite of the non hydrodynamic part of 
$B_2$ at $z=-1$. The results in [\rcite{CEM}] tell
us that the solution approaches, as $z^-\to\infty$ a function 
$q^- _2$ in  $\text{Null}\ {\cal L}^-$. Thus we set $b^-_2= h-q^- _2$, which will go
to zero at infinity exponentially in $z^-$ and will be the boundary
layer correction we are looking for. 

In conclusion, we have
$$f_2(\u x,v; t)=M\sum_{i=0}^4t_i^{(2)}(x,y,-1; t)\chi_i(v)+ 
b^+ _2(x,y, 2\e^{-1}; t) -q^- _2,\phantom{..} z=-1, \phantom{..} v_z>0, \phantom{..}
t> 0$$ 
Since the coefficients $t_i^{(2)}$ can be chosen arbitrarily on the boundaries  we
use them  to compensate $ q^- _2$. To satisfy the impermeability
conditions we have to choose $t^{(2)}_3=0$ on the boundaries.
The coefficients of the hydrodynamic part of $B_2$  will, for $i\ne 0$, 
be determined  by  the compatibility condition for $n=4$. These are
 time-dependent  non-homogeneous  Stokes equations
(linear second order differential equations) in a slab,
together  with the b. c.
$t^{(2)}_i=q^- _{2 i}, \  i=1, 2, 4$.  Then
$t^{(2)}_0$ is found up to a constant that is chosen so that the total
mass associated to $f_2$ vanishes. Finally we get
$$f_2(x,y,\pm 1, {v_z \gtrless 0}; t )=\a^{\pm}_2 M_{\pm} + 
\ga^{\pm}_{2,\e},\phantom{...}\a^{\pm}_2 =t^{(2)}_0(x,y,\pm 1)- 
q^{(\pm)}_2(0)$$
 
Iterating this procedure it is possible to find all $f_n$. To prove that the terms in the
expansion have the right properties we use the results in [\rcite{CEM}] 
for the solutions of the Milne problem with a force, that we state below 
\vskip.2cm
Let $F(z)=-\nabla V(z)$ be a force vanishing at  infinity such that $V(x)$ and its
derivative are bounded. Define 
$$\tilde M= e^{-V(z)} M,\qquad \tilde L f=e^{-V(z)} {1\over\sqrt M}2Q(\sqrt M
f,M).$$
Consider the following Milne problem:
$$
v_z {\pt f \over \pt z}+ F(z){\pt f \over \pt v_z}=\tilde Lf  + s( z,
v),\qquad 0< z<+\infty
\Eq(3.28)
$$
$$
f(0,v)=h(v),\qquad v_z>0
\Eq(3.29)
$$
$$
\lim_{z\to + \infty}f(z,v_z)=l<+\infty
\Eq(3.30)
$$
$$\int dv v_z  \sqrt{  M} f=0\Eq(3.31)$$
$$\int dv  \sqrt  M s=0\Eq(3.31.1)$$
\vskip.2cm
In [\rcite{CEM}]  it is proven the following
\goodbreak
\noindent{\bf Theorem 3.1}
\vskip.2cm
\noindent{\it 
\item{1)} Suppose that for $r>3$ and some $\sigma' >0$ 
there are finite constants $c_1$ and $c_2$ such that
$$\sup_{v \in \Bbb R^3} (1+ |v|)^r |h (v)|<c_1 $$
$$\sup_{z\in \Bbb R^{+}} e^{\sigma' z}\sup _{v\in
\Bbb R^3}(1+|v|)^r |s(z,v)|<c_2
\Eq(3.33)
$$
Then there is a unique solution $f\in L_\infty (\Bbb R^+
\times \Bbb R^3)$   to the Milne problem \equ(3.28)--\equ(3.31).
Moreover there exist constants $c$ and $c'$ such that
$f$ verifies the conditions:

$$f_\infty \in\hbox{Null}\/ \tilde L 
\Eq(3.34)$$
$$\sup_{z \in \Bbb R^{+}}e^{\sigma z}\sup_{v \in
\Bbb R^3} (1+|v|)^r |
(f(z,v)-f_\infty (v))|<c 
\Eq(3.35)$$ for
any $\sigma <c'$.

\item{2)} Suppose that for fixed $r>3$, $\ell\ge1$ and some $\b>0$ 
$$\sup_{v \in \Bbb R^3} (1+|v|)^r 
\Big|{\pt^\ell h \over \pt{v_i^\ell}}\Big|+
\sup_{z\in \Bbb R^{+}}e^{\b z}\sup_{v\in
\Bbb R^3} (1+|v|)^r  \Big|{\pt ^\ell s
\over \pt{v_i ^\ell}}\Big|<c_\ell
\Eq(3.36) $$ 
for some constant $c_\ell$ and $i\ne 3$. 
Then there are finite
constants $c$ and $c'_\ell$ such that 
$$\sup_{z\in \Bbb R^{+}}e^{\b z}\sup_{v \in
\Bbb R^3} (1+|v|)^r  \Big|\big[{\pt
^\ell f \over \pt {v_i ^\ell}}-{\pt^\ell f_\infty \over
\pt {v_i ^\ell}}\big]\Big |< c'_\ell
\Eq(3.37)$$
for any $\s<c$.

\item{3)} If $\gamma:=\sup_{z\in
(0,+\infty)} [|F'|+|F|]$ exists  and is finite then for any
$\delta>0$ and for $\gamma$ sufficiently small there exists a finite constant
$C_\delta$ such that

$$\sup_{v\in R^d}\sup_{z\in (\delta, +\infty)} (1+|v|)^r \Big|
{\pt\over\pt z}f\Big |\le C_\delta,\qquad \sup_{v\in R^d}\sup_{z\in [\delta,
+\infty)} (1+|v|)^r \Big|[ {\pt f\over\pt v_z}-{\pt
f_\infty\over\pt v_z}]\Big|\le C_\delta
\Eq(3.38)
$$
}

\vskip.7cm
\numsec=4
\numfor=1
{\bf 4. Results in the time-dependent case.}
\vskip.5cm
The main properties of the $f_n$'s are summarized in Theorem 4.1 below.

\noindent {\bf Theorem 4.1}

\noindent {\it  
Suppose that there is $\bar t>0$ such that $p(t)$, $\theta(t)$ and $u(t)$
are smooth solutions of OBE, with
$||\nabla u(t)||_{H_s} +||\nabla \theta(t)||_{H_s} \le q$ for
sufficiently large $s$ and $0<t\le \bar t$. 
Then it is possible to determine functions $f_n$, $n=2,\dots,7$
satisfying, for $0<t\le \bar t$, equation \equ(3.12) and the conditions
$$f_n({\u x},v; 0)=f_n^0, \phantom{..}\quad \phantom{...}
f_n(x,y,\pm 1,v_x,v_y, {v_z\gtrless 0}; t )=
\a^{\pm}_nM_{\pm},\phantom{...} t> 0,\Eq(3.39)$$
$$\langle A \rangle =0 \phantom{....}\Eq(3.40)$$
$$\int_{\Bbb R^3\times \Bbb T^2 \times[-1,1]}dv\/dx\/dy\/dz
f_n=0\Eq(3.41)$$ 
Moreover, for any  $\ell\ge 3$ there is a
constant $c$ such that:  
$$\sup_{0\le t\le \bar t}| f_n|_{\ell,h}<c\/q \Eq(3.42)
$$ 
$$\sup_{0\le t\le \bar t}| A|_{\ell,h}<c\/q, \Eq(3.43)$$ 
for $h\le 1/(4 T_-)$. Here
$$|f|_{\ell,h} = \sup_{{\u x}\in \Bbb R^2\times [-1,1]}\sup_{v\in\Bbb
R^3}(1+|v|)^\ell\exp[h v^2] |f(\u x,v; t)|\Eq(3.44) $$ 
}

{\bf Proof}

The proof is achieved by showing that every step of the procedure described in the
previous section is correct, namely that the conditions on the source and on the
boundary conditions for the Milne problems that we have to solve at each step are
satisfied. Moreover we need to check the solvability conditions for the Stokes
equations. 
\vskip.2cm
{\it Step 1 }

The first step is finding the boundary layer term $b^+_2$ solving for any $(x,y)\in
\Bbb T^2$ the Milne problem for 
$g_2=b^+_2/\sqrt {\tilde M_+}$:
$$
v_z{\pt\over\pt z^+} g_2 +\e^2 G^+{\pt g_2\over\pt v_z}={\cal L}^+ g_2,
\Eq(3.45)
$$
$$ 
\sqrt {\tilde M_+} g_2(x,y,0, v;t)=\bar B_2(x,y,1,v;t)-q_2^+(x,y,v;t),\quad v_z>0,
t>0,\qquad
\langle v_z  g 
\rangle=0,
\Eq(3.45.1)$$
where $\tilde M_+ = \exp [-V^+(z^+)]M_+$, $-\pt_{z^+}V^+=\e^2 G^+$, $\bar B_2$ is the
non-hydrodynamical part of
$B_2$ given by
$\bar B_2={\cal L} ^{-1}\Big[v\cdot \nabla f_1 +\u G\cdot\nabla_v M-Q(f_1,f_1)\Big] $. 
Finally $q_2^+(v;t)$  is the limit at infinity  of the solution $\tilde b^+_2$ of the
same Milne problem with boundary condition $\bar B_2$, as explained in the previous
section.

The force $G^+$ has been chosen smooth and vanishing as $z^+$ goes to $+\infty$ in 
such a way as to satisfy the assumptions on the force in Theorem 3.1. Furthermore, 
for $\e$ small the force term
(and its derivative) in \equ(3.45) is small. The boundary conditions verify
\equ(3.33) by the property of ${\cal L}^{-1}$ (see [\rcite{CIP}]).
Hence, by Theorem 3.1, 
$ {\tilde M_+}^{-1/2} b^+_2$ satisfy \equ(3.34)--\equ(3.38).

In the same way we construct $b^-_2$ imposing the boundary condition in $z^-=0$ given
by $\bar B_2(x,y,-1,v;t)$.
\vskip.2cm
{\it Step 2}

As explained above the coefficients $t^{(2)}_i, i=1,2,4$ of  the hydrodynamical part
of $B_2$  are determined by the compatibility condition for $n=4$
$$\big\langle \chi_i\big[ \pt_t B_2 +v\cdot \nabla B_3 +
G\cdot\nabla_vB_2\big]\big\rangle$$
where 
$$
B_3={\cal L} ^{-1}\Big[\pt_t f_1 +v\cdot \nabla B_2 +\u G\cdot\nabla_v
f_1-2Q(f_1,B_2)\Big] +M\sum_{i=0}^ 4\chi_i\  t^{(3)}_i({\u x}, t)
\Eq(3.46)
$$

Proceeding as in the determination of the Boussinesq equation, we find now a set of
three linear time-dependent non-homogeneous  Stokes equations 
for $t^{(2)}_i$. The non-homogeneous terms  depend on the third order spatial
derivatives of 
$f_1$.  
We note that the non linear terms in the hydrodynamic equations come from the
quadratic term
$Q(f_1,f_1)$ in
\equ(3.24), while in
\equ(3.46) appears a term linear in
$B_2$. General theorems for the Stokes equation assures the existence of a solution for 
 the  chosen boundary  and initial conditions. 
\vskip.2cm
{\it Step 3}

Once $B_2$ is completely determined, \equ(3.46) gives the non-hydrodynamical part of
$B_3$, $\bar B_3$. As before, we introduce the terms $b^\pm_3$ to compensate $\bar B_3$
on the boundaries $z=\pm 1$. The term $b_3^+$ is found as a solution of the Milne
problem for $g_3: \ \sqrt{\tilde M_+} g_3= b_3^+$
$$v_z{\pt g_3\over \pt z^{+}}+ \e^2 G^+\pt_{v_z} g_{3} ={\cal L}^+ g_3 +s(x,y,z^+,v;t)
\Eq(3.47)$$
with source
$$\sqrt {\tilde M_+} s(x,y,z^+,v; t)= \hat v\cdot\hat\nabla b^{+}_{2}+2Q(\Delta
M,b^+_{2})+ 2Q(f_1,b^{+} _2)
\Eq(3.48)$$
and with boundary condition
$$\sqrt {\tilde M_+} g_3(x,y,0,v;t)=\bar B_3(x,y,1,v;t)-q^+_3(v;t)\quad v_z>0, t>0 .
$$
We have to check that the source satisfies the conditions of Theorem 3.1. 

The condition \equ(3.31.1) is true due to the properties  of $Q$ and to \equ(3.31) for
$b^+_2$. The terms of the form $Q(f,g)$ are bounded by means of the Grad estimates 
$$|M^{-{1\over 2}} Q(f,g)|_{r-1}\le C ||M^{-{1\over 2}}f|_{r} ||M^{-{1\over 2}}g|_{r}
\Eq(3.49)$$
so that
$$|{\tilde M_-}^{-{1\over 2}}Q(f_1,b^{+} _2)|_{r-1}\le C|{ \tilde M_-}^{-{1\over 2}} f_1|_{r}
|{\tilde M_-}^{-{1\over 2}} b^+_2|_{r}$$
The second term in \equ(3.48) is bounded in the same way 
$$|{\tilde M_-}^{-{1\over 2}}Q(\Delta M,b^+_{2})|_{r-1}\le \Bigg|\e^{-1}{M-M_-\over
\sqrt {\tilde M_-}} \Bigg|_{r}|{\tilde M_-}^{-{1\over 2}}b^-_{2} |_{r}
$$
Since $T_-=T_+(1+2\e\l)$ we have that $|\e^{-1}{M-M_+\over \sqrt {\tilde M_-}}
|_{r-1}\le C|({v^2\over T_-}+1)\sqrt M|_{r-1}$.

The functions $f_1$ and $b^+_{2}$ have bounded norm, hence the third term in
\equ(3.48) is bounded.  To conclude the step we need to bound the first term. It is
bounded by using  the properties for the derivatives with respect  to $x$ and
$y$ of $b^+_2$ assured by Theorem 3.1.

{\it Step 4}

$B_3$ is constructed as in Step 2. Instead the Milne problem for $b^+_4$   has to be
discussed since in the source appear also derivatives of $b^+_2$ with respect $t$ and
$v_z$. We have for $\sqrt{\tilde M_+}  g_4= b^+_4$ that
$$
v_z{\pt g_4\over \pt z^{+}}+ \e^2 G^+{\pt\over\pt{v_z}}g_{4} ={\cal L}^+g_4 +s_4,
$$
$$\eqalign{
\sqrt{\tilde M_+} s_4(x,y,z^+,v;t)&= \partial_t  b^+_2  -(G^0+G^-){\pt\over\pt{v_z}} 
b^{+}_{2}+\hat v\cdot\hat\nabla b^{+}_{3} +2Q(\Delta M,b^+_{3})+\\
&\sum_{\st i,j\ge 1 \atop\st i+j=4 } \Big[2Q(B_i,b^{\pm} _j) + Q(b^{\pm}_i,
b^{\pm}_j) +Q(b^{\mp} _i,b^{\mp} _j)\Big] .}
\Eq(3.50)
$$

The first term in r.h.s. of \equ(3.50) is bounded by observing that the time
derivative of $b^-_2$ is a solution of the Milne problem we get by differentiating
\equ(3.45) with respect to time with boundary condition $\partial_t  b^-_2(0,v;t)$,
$v_z>0$, 
$t>0$. 

The second term in r.h.s. of \equ(3.50) is zero by construction for $0\le z^-\le
\delta$ so that
$$|{\tilde M_-}^{-1/2} (G^0+G^-) \nabla_{v_z} b^{+}_{2}|_{r-1}\le C \Big\{1+ 
\sup_{v\in R^d}\sup_{x\in (\delta, +\infty)} (1+|v|)^r \Big| [ {\pt g_{2}\over\pt
v_z}-{\pt {g_2}_\infty\over\pt v_z}]\Big|\Big\}$$
and it is bounded because of \equ(3.38), Theorem 3.1, that applies as shown in Step 1.

Finally the condition \equ(3.31.1) is satisfied since the velocity flux of the 
derivatives of $b^-_2$ with respect to time and velocity is zero.

\vskip.2cm

The terms $f_n$ with higher $n$ are constructed and  bounded in the same way. 
As a consequence, it is easy to see by using the preceding arguments that the term
$A$ in \equ(3.14) satisfies the bound  \equ(3.43) and \equ(3.40). This concludes
the proof of Theorem 4.1.

\vskip1cm
To complete the construction of a solution to the BE, we have to show that the
remainder is bounded in  norm $|\cdot|_{\ell,h}$ defined in \equ(3.44). The remainder
has to satisfy \equ(3.14) and the conditions \equ(3.19) and \equ(3.20) with $\langle
\gamma^\pm_{n,\e} v_z\rangle=0$.   Moreover $R$ has to satisfy 
$$R( {\u x},v; 0)=0.,\qquad \langle v_z R \rangle=0 \ \text{ in } z=\pm 1\Eq(4.10)$$
which implies  $\a^{\pm}_R= \pm \int_{v_z \gtrless 0} v_z R(\pm 1,v; t)dv$. To
construct the solution of \equ(3.14),\equ(3.19), \equ(3.20) and \equ(4.10) we first
deal with the following linear initial boundary value problem: given $D$ on
$\Bbb T^2\times [-1,1] \times \Bbb R^3$,  find $R$  such that
$$\pt_t R+ \e^{-1}v\cdot\nabla R +\u G\cdot\nabla_v R=
\e^{-2} {\cal L} R +\e^{-1}{\cal L}^1 R +{\cal L}^2 R+ D, \Eq(4.20)$$
 with the same initial and boundary  conditions as before. 

Once one obtains good estimates for the solution of this linear problem,  the non
linear problem is solved by simple Banach fixed point arguments, for small $\e$. 
This allows to conclude the existence of the solution $f^\e$ and its convergence to
the solution of the OBE.

\vskip.4cm
\noindent$\underline{\hbox{\rm Solution of Linear problem.}}$
\vskip.3cm

We consider the linear problem \equ(4.20) with a given $D$ satisfying 
$\langle v_z D\rangle=0$.
Put $R=\sqrt{\tilde M} \Phi$. Therefore the equation for $\Phi$ is 
$$\pt_t \Phi+ \e^{-1}v\cdot\nabla \Phi +\u G\cdot\nabla_v \Phi=
\e^{-2} \tilde L \Phi +\e^{-1}\tilde  L^1 \Phi +\tilde  L^2 \Phi+ \tilde D, 
\Eq(4.21)$$ 
where $\tilde D= \tilde M^{-1/2} D$, $\tilde M=M \exp[\e G(z+1)/T_-]$
and 
$$\tilde L^jf= \tilde M^{-1/2} {\cal L}^j \tilde M^{1/2} f,\quad j=1,2 $$
The boundary and initial conditions are
$$ \Phi(\u x, v;0)=0;\qquad \Phi(x,y,\pm1,v; t)= \a^\pm_R \overline
M_\pm(v)\tilde M^{-1/2} +\zeta^\pm,
\qquad v_z \lessgtr 0,\phantom{..}t> 0
\Eq(4.21.1)$$
where
$$\zeta^\pm:=-{\tilde M}^{-1/2}\sum_{n=2}^7  \e ^{n-3} \ga^\pm _{n,\e}, \qquad
\a^\pm_R=\pm
\int_{v_z\gtrless 0}v_z\sqrt{\tilde M}\Phi(x,y,\pm 1,v;t) dv
\Eq(4.21.2)$$

\vskip.2cm
\noindent Introduce the $L_2$ norm as
$$\pa f\pa =\Big\{\int _{\Omega\times \Bbb R^3} d\u x dv |f(\u x,v;t)|^2\Big\}^{1/2}$$
We now give first the $ L_2$ bound for $\Phi$ and then we provide the $L_\infty$ bound.
\vskip.3cm
 
\goodbreak

{\bf Theorem 4.2}

\vskip.2cm
{\it
The solution of the linear problem \equ(4.21), \equ(4.21.1) satisfy the bound
$$||\Phi||\le C\sup_{t\in(0,\bar t]} ||\tilde D||$$
}

{\it Proof}

Multiplying \equ(4.21) by $\Phi$ and integrating on $\Omega\times \Bbb R^3$ we have 
$$\eqalign{
{1\over 2} \pt_t \pa \Phi\pa^2 &= \e^{-2}\int_\Omega d\u x\int_{R^3} dv     
\Big[\Phi \tilde L \Phi +\e\Phi {\tilde  L}^1\Phi + \e^2\Phi {\tilde  L}^2\Phi +\Phi
\tilde D\Big] \\ &  +{1\over 2 \e}\int dx dy\Big[ \langle v_z\Phi ^2\rangle
(x,y,-1;t) - \langle v_z\Phi ^2\rangle(x,y,1;t)\Big]
}
\Eq(4.22)
$$
To bound the boundary terms in the second line of \equ(4.22), we proceed as follows.
We consider first the more difficult term
$$\langle v_z \Phi^2  \rangle(x,y,1;t)=- ({\a_R^+})^2\int_{v_z<0}dv| v_z|  {\overline
M}_+^2 \tilde M^{-1} + \int_{v_z>0}dv v_z \Phi(x,y,1,v; t)^2 .\Eq(4.23)$$ 
 
\noindent By the Schwartz inequality, 

$$\eqalign{
({\a_R^+)}^2 &= \Big(\int_{v_z>0}d\/v  |v_z |^{1\over 2}  |v_z|
^{1\over 2} \tilde M^{1\over 2} \Phi(x,y,1,v; t)\Big )^2 \\
&\le  \int_{v_z>0}dv |v_z| \Phi^2 (x,y,1,v; t)
\int_{v_z>0}dv |v_z| \tilde M
}
\Eq(4.24)$$ 
Using the relation between $\tilde M$ in $z=+1$ and $\overline M_+$, the 
normalization of $\overline M_+$, and the relation $T_+=T_-(1-2\e\l)$, we get, after a
straightforward computation, 
$$\eqalign{
&\int_{v_z>0}dv |v_z| \tilde M\int_{v_z>0}dv v_z {\overline M}_+^2
\tilde M^{-1}=\Big({T_-\over2\pi}\Big)^{1/2} \int_{v_z>0}dv v_z {\overline M}_+^2
M^{-1}=\\&
\Big({T_-\over2\pi}\Big)^{1/2}{(2\pi T_-)^{3/2}\over (2\pi T_+^2)^2}2\pi
\left({T_+T_-\over 2T_--T_+}\right)^2={(1-2\e\l)^2\over (1-2\e\l)^2(1+2\e\l)^2}\\&=
(1-\e^2\l^2)^{-2}\le 1+ C\e^2
}
\Eq(4.25)
$$
Using this bound and \equ(4.24) in \equ(4.23) we find
$$- {1\over\e} \langle v_z \Phi^2 (x,y,1,v; t)
\rangle\le C\e\int_{v_z>0}dv |v_z| \Phi^2 (x,y,1,v; t)\Eq(4.25.2).$$           

\noindent The same argument shows that $\langle v_z \Phi^2
(x,y,- 1,v; t) \rangle$  is non positive because
$$\int_{v_z<0}dv |v_z| \tilde M\int_{v_z>0}dv v_z {\overline M}_-^2
\tilde M^{-1}=1.$$

In Appendix it is proved the following bound for the r.h.s. of \equ(4.25.2): let $t_1$
be any time in $(0.\bar t]$. Then

$$\eqalign{
&\e\int_0^{t_1}dt\int_{\Bbb T^2} dxdy \int_{v_z>0}dv |v_z| \Phi^2 (x,y,1,v; t)\\&\le
\int_0^{t_1}dt\pa \Phi(\,\cdot\,,t)\pa^2 +\e^4\sup_{0<t\le t_1}\pa\tilde
D(\,\cdot\,,t)\pa^2+\sup_{0<t\le t_1} |\zeta(\,\cdot\,,t)|^2 }
\Eq(4.24.a)$$
where $|\zeta|=|\zeta^+|+|\zeta^-|$.

\vskip.3cm
We recall the following crucial properties of the linearized Boltzmann operator
$\tilde L$ (see for example [\rcite{CIP}]): 
\item{1)} $$\tilde L= -\nu +K,$$
with $K$ an integral operator and $\nu$ a smooth function. Moreover, for hard spheres,
there are two constants $\nu_0$ and $\nu_1$ such that
$$ \nu_0(1+|v|)\le\nu(x,v)\le \nu_1(1+|v|);$$
\item{2)} There is a constant $C>0$ such that
$$\langle \Phi \tilde L\Phi\rangle\le -C\langle \nu\bar\Phi^2\rangle.\Eq(LB)$$
with the usual decomposition $\Phi=\bar \Phi
+\hat\Phi$, with $\bar \Phi$ and $\hat\Phi$ the non-hydrodynamical and the
hydrodynamical part of $\Phi$ respectively.
\vskip.2cm
To estimate the operators ${\tilde L}^1$ and
${\tilde L}^2$ we will use the following estimate on the
collision operator $Q$ (see for example [\rcite{GPS}]): for any Maxwellian
$M$ and for any $y\in[-1,1]$  
$$\int_{\Bbb R^3}\ dv  
{|Q (\sqrt M f,\sqrt M g)|^2\over \nu M}
\le \int_{\Bbb R^3}\ dv \nu |f|^2\ \int_{\Bbb R^3}\ dv  \nu|g|^2
\Eq(4.24.1)$$
This inequality and the bounds on the $f_n$'s imply the following bounds:
$$\int d\/x\/d\/v\Phi {\tilde  L}^1\Phi\le C\pa \sqrt\nu \bar 
\Phi\pa \ \pa 
\Phi\pa \pa M^{-1/2}f_1\pa,\Eq(BL1)$$
$$\int d\/x\/d\/v\Phi {\tilde  L}^2\Phi\le C\pa \sqrt\nu
\bar\Phi\pa \ \pa 
\Phi\pa \pa  M^{-1/2}\sum_{n=2}^7
f_n\pa.\Eq(BL2)$$
Note that the presence of the product $\pa\sqrt\nu\bar\Phi\pa\/\/ \pa\Phi\pa$
depends on from the fact that $\tilde L^1$ and $\tilde L^2$ are both orthogonal to
the collision invariants.

We integrate \equ(4.22) on time between $0$ and $ t_1$, and recall that
$\Phi(x,v,0)=0$. With the notation $\Phi_t(\,\cdot\,)=\Phi(\,\cdot\,,t)$, we get

$$\eqalign{&{1\over 2}  \pa \Phi_{t_1} \pa^2 \le
C\int_0^{t_1}d\/t\Bigg\{-\e^{-2}\pa 
\sqrt\nu\bar \Phi_t\pa^2\\& +\pa \sqrt\nu \bar \Phi_t\pa \ \pa 
\Phi_t\pa
\left[\e^{-1}\pa M^{-1/2}f_1\pa + \pa  M^{-1/2}\sum_{n=2}^7
f_n\pa\right]+\pa\tilde
D_t\pa^2\Bigg\}\\&+C\int_0^{t_1}d\/t\pa  \Phi_t\pa^2
+\e^4\sup_{0<t\le t_1}\pa\tilde D(\,\cdot\,,t)\pa^2+\sup_{0<t\le
t_1} |\zeta(\,\cdot\,,t)|^2 }\Eq(4.24.3)
$$
The last line in \equ(4.24.3) derives from the bound \equ(4.24.a). The first 
term in the second line is due to the bounds \equ(BL1) and \equ(BL2).  
Moreover $ \pa M^{-1/2} f_1\pa\le  C$  by the regularity   of the solutions of the
macroscopic equations for $0<t<\bar t$ and $\pa \tilde M^{-1/2}\sum_{n=2}^7 f_n\pa\le
C$ by Theorem 4.1.

The following elementary inequality 
$$   -{1\over\e^{2}} x^2  +(c_1\e^{-1}+c_2)x y \le (c_1\e+c_2)^2y^2/4$$
is valid for any positive $\e$, $x$, $y$.
We apply it with $x=||\sqrt{\nu}\bar\Phi||$, $y=||\Phi||$ and suitable
constants $c_1$ and $c_2$. We get
$$\eqalign{&\pa \Phi(\,\cdot\,,t_1)\pa^2 \le\int_0^{t_1}d\/t 
C\Bigg[\pa\Phi(\,\cdot\,,t)\pa^2 +\pa\tilde D(\,\cdot\,,t) \pa^2
\Bigg] \\&+\e^4\sup_{0<t\le t_1}\pa\tilde D(\,\cdot\,,t)
\pa^2+\sup_{0<t\le t_1} |\zeta(\,\cdot\,,t)|^2}$$
In conclusion, by the use of the Gronwall lemma, for $\e$ sufficiently small, we get:
$$\sup_{0\le t\le \bar t} \pa \Phi(\,\cdot\,,t)\pa\le C_{\bar t} \sup_{t\in(0,\bar t]}
\pa \tilde D(\,\cdot\,,t) \pa\ + \sup_{t\in(0,\bar t]}
|\zeta(\,\cdot\,,t)| \Eq(4.25.1)$$
\vskip.3cm

\vskip.4cm
{\it $L_\infty$ bound.}
\vskip.3cm
Let us give some notation:

$$\pt \Lambda_\pm:= \Big\{ \u x=(x,y,z): (x,y) \in \Bbb T^2,
z=\pm 1
\Big\}$$

$$S:=\Big\{(t, \u x, v): t\in(0,\bar t], (\u x, v)\in \Omega\Big\}\Eq(4.26)$$
$$\pt S:=\Big\{(t, \u x, v): t=0, (\u x, v)\in
\Omega\Big\}\cup
\Big\{(t, \u x, v): t \in(0,\bar t], \u x\in\pt\Lambda_\pm, 
v_z\lessgtr 0\Big\}.\Eq(4.27)$$
$\Lambda $ and $\Omega$ are defined in \equ(3.0).

We call $\f_t(x,v)$ the characteristics of the equation 
$$\pt_t f +{1\over \e}v\cdot \nabla_x f +\u G\cdot\nabla_{v}f=0$$
given by
$$\f_t(x,v)=(\u x+{v\over \e}t+{1\over 2\e}\u Gt^2,v+ \u
Gt).\Eq(4.28)$$

and define
$$t^-=\inf\{t\ge 0 : (t, \f_t(x,v))\in S\};\qquad
t^+=\sup\{t^-<t<\bar t: (s, \f_s(x,v))\in S\,\, \forall s\le
t\}\Eq(4.29)$$ 
Given smooth functions $H$, $\tilde\nu$ on $S$ and $h$ in $\pt S$, consider the
initial boundary value problem
$$\pt_t f +{1\over \e}v\cdot \nabla_x f +\u G\cdot \nabla_{v_z}f-{1\over
\e^2}\tilde\nu f=H
\Eq(4.30)$$
$$f(t,\u x,v)=h(t,\u x,v), \qquad (t,\u x,v)\in \pt S.\Eq(4.30.1)$$
The solution of this problem is written as
  $$\eqalign{
  f(t,x,v)&= h\big(t^-, \f_{t^--t}(x,v)\big)\exp\Big\{-\int_{t^-}^t ds{1 \over
  \e^2}\tilde\nu\big(\f_{s-t}(x,v)\big)\Big\}\\ & +\int_{t^-}^t ds
  H\big(s,\f_{s-t}(x,v)\big)\exp\Big\{-\int_{s}^t ds'{1\over \e^2}
  \tilde\nu\big(\f_{s'-t}(x,v)\big)\Big\}. }\Eq(4.31)
  $$
We introduce the  norm
$$||f||_{p,q}:=\Bigg[\int_{\Bbb T^2\times [-1,1]} d\u x \Big[\int_{\Bbb R^3} dv
|f|^q\Big]^{p\over q}\Bigg]^{1\over p}\Eq(4.32)$$
and denote by $L^{q,p}$ the corresponding space. We define the operator
$N_{_s}^t$ as
  $$N_{s}^tf= f\big(s,\f_{s-t}(x,v)\big)\exp\Big\{-\int_{s}^{t}ds'{1\over\e^2}
  \tilde\nu\big(\f_{s-t}(x,v)\big)\Big\}.\Eq(4.32.2)
  $$ 
We will also omit the apex
$t$ when there is no risk of confusion.

We assume that $\tilde\nu$  corresponds to the collision rate for hard spheres,
i.e.
$$\nu_0(1+|v|)\le \tilde\nu\le \nu_1 (1+|v|),\Eq(nupos)$$
Then $N_s^t$ satisfies the estimate
$$||N_s^t f||_{p,q}\le C  \exp\{-{\nu_0(t-s)\over\e^2}\}||{f}||_{p,q}\Eq(4.32.3)$$

The following Lemma allows to bound the $L^{p,q}$ norm  of $N_sf$  in
terms of the $L^{q,p}$ norm of $f$. It is a generalization to the case with a 
constant force of the theorem of Ukai and Asano [\rcite{UA}].

\vskip.2cm

      \noindent{\bf Lemma 4.3}
      
      {\it 
      Let $1\le q,p\le+\infty$ and
        $\a={1\over q}-{1\over
      p}$. Then
      $$||N_s^t f||_{p,q}\le \Big({\e\over t-s}
      \Big)^{d\a}\exp\{-{\nu_0(t-s)\over\e^2}\}||{f}||_{q,p}.
      \Eq(4.32.4)$$  }
\vfill\eject
\noindent{\it Proof.}
\vskip.1cm
Consider the following change of variables $v\to \u y$:
$$\u y=\u x+{1\over \e}v\tau +{1\over 2\e}\u G\tau^2,\quad \hbox{
or  }\   v+{1\over 2}\u G\tau=\e {\u y-\u x\over
\tau}\Eq(change)$$ where we have put $\tau=s-t$. We have under
the change
\equ(change) 
$$
||N_s^t f||_{p,q}\le \Big({\e\over \tau}\Big)^{{d\over
q}}\exp\{-{\nu_0\tau\over\e^2}\}\Bigg[\int d\u x\Big[\int d\u y \Big| {f}(\tau
+t, \u y,\e {\u y-\u x\over \tau}+{1\over 2}G\tau)\Big|^q\Big]^{p\over
q}\Bigg]^{1\over p}
$$ 
With one more change of variables 
$$\u x\to w=\e {\u y-\u x\over\tau}+{1\over 2}G\tau,$$ 
we get
$$
||N_s^t f||_{p,q}\le \Big({\e\over \tau}\Big)^{{d\over q}}\Big({\e\over\tau}
\Big)^{-{d\over p}}\exp\{{\nu_0\tau\over\e^2}\}||f||_{p,q}$$
Hence the result.

\vskip.2cm
This Lemma will be used also for $p=+\infty$ and $q>d$ to control
the
$(\infty,q)$-norm of $N^t_s f$ in terms of $(q,\infty)$-norm of
the  solution of \equ(4.31).
\vskip.2cm
We write  \equ(4.21), \equ(4.21.1) in the form \equ(4.30), 
\equ(4.30.1) with
$$H=
\e^{-2}\tilde K\Phi +\e^{-1}\tilde K^1\Phi +\tilde K^2 \Phi+\tilde D:=\e^{-2}
\tilde K \Phi+H', \Eq(4.34)$$ 
$\tilde L$ is the linear Boltzmann operator such that
$$\tilde { L}:=\tilde K -\nu,$$
$$\tilde L^i:=\tilde K^i-\nu^i,\qquad i=1,2.$$ 
Here the $\nu^i$ are defined analogously to $\nu$ as  
$$\eqalign{
\nu^1(\u x, v)&=M^{-1/2}\int dv'd\omega(v'-v)\cdot \omega 
f_1\sqrt{\tilde M},\\
\nu^2(\u x, v)&=M^{-1/2}\int dv'd\omega(v'-v)\cdot \omega
\sum_{n=2}^7 f_n\sqrt{\tilde M}.
}
$$
Finally we set $\bar\nu= \tilde \nu +\e \tilde \nu^1 +\e^2\tilde \nu^2$ and it
is immediate to check that $\bar \nu$ satisfies the assumption \equ(nupos).

\goodbreak
We have the following
   \vskip.2cm
   \noindent{\bf Theorem 4.4}
   \vskip.2cm
   {\it 
   \noindent Let $\Phi$ be the solution of the problem \equ(4.30),
   \equ(4.30.1), with $\tilde \nu$, $H$ and $h$ given as before. 
   Then, for any $q>d$, ${1\over 2}-{1\over d}<{1\over q}$,
   $$\eqalign{
   \sup_{0\le t\le\bar t} \pa\Phi\pa_{\infty,q}&\le C\e^{-d/2} \sup_{0\le
   t\le\bar t} \pa\tilde D \pa_{2,2}+ C\e^{-d/2}\sup_{0\le t\le\bar
   t}\pa\zeta\pa_{2,2}\\&+C\e^2 \sup_{0\le t\le\bar t}\pa H'\pa_{\infty,q}+
   C\e^2 \sup_{0\le t\le\bar t}\pa\zeta\pa_{\infty,q}.}
   \Eq(4.32.1)$$}
   \vskip.2cm
\noindent{\it Proof.}
\vskip.1cm
By \equ(4.31) we get 
$$\pa\Phi\pa_{\infty,q}\le  \pa N_{t^-}^t h\pa_{\infty,q}+\int _{t^-}^t ds\Pa
N_s^t H'\Pa_{\infty,q}+ \e^{-2}\int _{t^-}^t ds \pa N_{s}^t\tilde K
\Phi\pa_{\infty,q}\Eq(a.1)$$ 
In the last term  we substitute to $\Phi$ its expression \equ(4.31) so that 
$$\eqalign{
&\pa N_s^t\tilde K \Phi\pa_{\infty,q}\\&\le\pa N_{s}^t\tilde K N_{s^-}^s
h\pa _{\infty,q}+\int _{s^-}^s ds'\Bigg[
\pa N_s^t\tilde K N_{s'}^s H'\pa _{\infty,q} +\e^{-2}\pa N_s^t\tilde K N_{s'}^s \tilde
K \Phi\pa _{\infty,q}\Bigg ].}\Eq(a.2)$$
The boundary term will be discussed separately. To estimate the terms
containing $\tilde K $ we use the following Lemma whose proof is given in [\rcite{UA}]
\vskip.2cm
\goodbreak
    \noindent {\bf Lemma 4.5}
    \vskip.2cm
    \noindent {\it 
    Let $L^s_{\gamma}$ the spaces  of functions $f(v)$ such that
    $$\int_{\Bbb R^3}dv |f(v)|^s(1+|v|)^\gamma<\infty$$ 
    with $\gamma\in \Bbb R$. 
    Let $1\le s\le r\le \infty$ and $\eta_0=1-{1\over s}+{1\over r}$. 
    Then  
    the operator $K$ maps $L^s_\gamma$ into $L^r_{\gamma +\eta}$  if  
    ${1\over s}-{1\over r}<{2\over d}$ and $\eta\le \eta_0$.
    }
    \vskip.4cm
We have by \equ(4.32.3) and by Lemma 4.5
$$\eqalign{
\pa \Phi\pa _{\infty,q}&\le  \sup_{0\le t\le\bar t}\pa N_{t_-}^t h\pa
_{\infty,q}+C \sup_{0\le t\le\bar t}\Pa { H'}
\Pa_{\infty,q} \int_{t_-}^t ds \exp\{-{\nu_0(t-s)\over\e^2}\}\\
&
+\e^{-4}
\int _{t^-}^t ds\int _{s^-}^s ds'\pa N_{s}^t\tilde K N_{s '}^s \tilde K \Phi\pa _{\infty,q}
}
\Eq(a.3)$$
Using the bound $\int_{t_-}^td\/s \exp[-\e^{-2}\nu_0(t-s)]\le
C\e^2$, the estimate for the second term in the r.h.s.
of \equ(a.3) follows.  We bound the last one  by using  Lemmas
4.3 and 4.5  as follows. 

$$
\eqalign{
\pa N_s^t\tilde K N_{s '}^s \tilde K
\Phi\pa _{\infty,q}&\le C\Big( {\e\over (t-s)}\Big)^{d/q}
\exp\{-{\nu_0\over\e^2}(t-s)\}\pa \tilde K N_{s'}^s
\tilde K\Phi\pa_{q,\infty}\\ &\le C\Big( {\e\over (t-s)}\Big)
^{d/q}\exp\{-{\nu_0\over\e^2}(t-s)\}\pa  N_{s'}^s \tilde K
\Phi\pa _{q,2}\\
&\le C\Big({\e\over (t-s)}\Big)^{d/q}\Big({\e\over (s-s')}\Big)^{d\beta}\exp\{-{\nu_0\over\e^2}[(t-s)+(s-s')]\}\pa  
\tilde K
\Phi\pa _{2,q}\\
&\le C\Big({\e\over (t-s)}\Big)^{d/q}\Big({\e\over (s-s')}\Big)^{d\beta} \exp\{-{\nu_0\over\e^2}[(t-s)+(s-s')]\}\pa  
\Phi\pa _{2,2}
}
\Eq(4.40)
$$

To get the first inequality we have used \equ(4.32.3) and the Lemma 4.3 which
hold for $q>d$. The second step is based on the Lemma 4.5 to replace the
$L^\infty$ norm on the velocity with the $L^2$ norm, by taking
$r=+\infty$, $s=2$, $\gamma=0$. Then the Lemma 4.3 allows to exchange the
exponents for space and velocity introducing a factor $(\e/(s-s'))^{d\b}
\exp[-\nu_0(s-s')/\e^2]$, with $\beta={1\over 2}-{1\over q}$. Finally the Lemma
4.5 again, with $r=q$, $s=2$ and $\gamma=0$,  gives the bound in terms of the
$L^2$ norm in space and velocity. 

To get convergence of the $s'$-integral we need  
$0<\beta<{1\over d}$, so we have to choose $q>d$ , ${1\over
2}-{1\over d}<{1\over q}$. From the time integrations we get a
factor
$\e^{4-2(\mu+\mu')}$ with $\mu=d/q, \mu'=d\beta$. Combined with the prefactor
$\e^{(\mu+\mu')}$ it produces $\e^{4-(\mu+\mu')}$. But $\mu+\mu'=d/2$ so that
the factor we gain is
$\e^{4-d/2}$. In conclusion
$$\eqalign{
\sup_{0\le t\le\bar t} \pa \Phi\pa _{\infty,q}&\le C\sup_{0\le t\le\bar t}\pa
N_{t_-}h\pa _{\infty,q}+C\e^2 \sup_{0\le t\le\bar t}\pa   H'\pa _{\infty,q}\\
&+C\e^{-d/2}\sup_{0\le t\le\bar t} \pa 
\Phi\pa_{2,2}
}
\Eq(4.41)$$

Now we bound the boundary term. We recall that its  meaning is as follows:
define 
$$\tau_-=\inf\{\tau>0:\f_{-\tau}(\u x,v; t)\in\pt S\}$$
If $\tau_-=t$ then $h=\Phi(\u x,v; 0)=0$; if $\tau_-< t$ then 
$h=\Phi(x,y,\pm 1, v;t)$, in correspondence of $v_z\lessgtr 0$
and $(\f_{-\tau}(\u x,v; t)\big)_z=\pm 1$.
The boundary conditions on $\Phi$ are given by \equ(4.21.1)  with
$$\a_R^{\pm}=\pm\int _{v_z\gtrless 0} v_z\sqrt{\tilde M}\Phi(x,y,\pm1,v; t)
dv$$

Equation \equ(4.31) allows to express $\a_R^+$  in terms of
$\Phi(x,y,-1,v,t), v_z<0$ and $H$. In fact by \equ(4.31)

$$\eqalign{
\Phi(x,y,1,v;t)&= h\big(t^-, \f_{t^--t}(x,y,1,v)\big)\exp\Big\{-\int_{t^-}^t 
ds{1 \over\e^2}\nu\big(\f_{s-t}(x,y,1,v)\big)\Big\}\\ & +\int_{t^-}^t ds
H\big(s,\f_{s-t}(x,y,1,v)\big)\exp\Big\{-\int_{s}^tds'{1\over \e^2}
\nu\big(\f_{s'-t}(x,y,1,v)\big)\Big\}. }
\Eq(h.1)
$$

We remark that the characteristic $\f$ which is on the boundary $z=1$ with 
velocity $v$ such that $v_z>0$ at time $t^- -t$ had to start  either from some
point in the bulk or from the boundary $z=-1$. In the first case  we have $h=0$
in \equ(h.1) otherwise we get  $h=\Phi(x,y,- 1, v;t)$ for $v_z>0 $. In the
latter case it is shown in the Appendix that the integral $\int_{t^-}^t
ds\nu\big(\f_{s-t}(x,y,-1,v)\big)$ is bounded from below by $C\e$.  As
a consequence, we can estimate the exponential in the first row of \equ(h.1) as
$e^{-C/\e}$. 
 
We exploit the same argument to deal with $\a^-_R$ and use 
\equ(4.31) to represent $\Phi$ on the boundary $z=-1$. There is
a difference with respect to the previous case: due to the
presence of the force the characteristic $\f$ which is on the
boundary $z=-1$ with velocity $v, v_z<0$ at time $t^- -t$ can
start from the bulk, from the boundary $z=1$ with negative $v_z$
or from the boundary $z=-1$ with positive $v_z$. In the Appendix
it is shown that in the latter case  the exponential factor in
the first row of \equ(h.1) allows to gain a factor $C\e^2$.

We discuss explicitely the bound for $\a^-_R$. The case of $\a^+_R$ is dealt 
with in the same way. By using  the representation
\equ(4.31) we get

$$\eqalign{ 
\sup_{(x,y)\in \Bbb T^2} &|\a_R^-(x,y,-1;t)|=\sup_{(x,y)\in \Bbb T^2}\Big|\int
_{v_z> 0}  v_z\sqrt {\tilde M}\Phi(x,y,-1,v; t) dv \Big|
\\&
\le C\sup_{(x,y)\in \Bbb T^2}\Big[\int_{v_z>0}dv|\Phi|^q(x,y,-1,v;
t)\Big]^{1/q}\le\\&
C\sup_{(x,y)\in \Bbb T^2}[e^{-{C\over\e}}|\a^+_R|+\e^2|\a^-_R|] +\e^2 \pa
\zeta\pa _q +\int_{t^-}^t ds\Pa N_s^t H\Pa_{\infty,q}\le\\&
C\e^2\pa\Phi\Pa_{\infty,q}+ \int _{t^-}^t ds\Pa N_s^t H'\Pa_{\infty,q}+
\e^{-2}\int _{t^-}^t ds \pa N_{s}^t\tilde K \Phi\pa _{\infty,q}
}
\Eq(h.1.1)
$$
The first bound is obtained from the inequality

$$\eqalign{
\sup_{(x,y)\in \Bbb T^2}&|\a^\pm_R|\le \sup_{\u x\in\Bbb T^2\times[-1,1]}\int
_{v_z\gtrless 0} v_z\sqrt {\tilde M}|\Phi(\u x,v; t)| dv \\&\le 
C\sup_{\u x\in\Bbb T^2\times[-1,1]}\Big[\int_{v_z>0}dv|\Phi|^q(\u x,v;
t)\Big]^{1/q}=\pa \Phi\pa _{\infty,q} }$$
As explained before, to get a bound of the last term in \equ(h.1.1) in terms 
of the $L_2$ norm we need to iterate the procedure and use again the
representation formula 
\equ(4.31) in the last term
$$\eqalign{
\e^{-2}\int _{t^-}^t ds &\pa N_{s}^t\tilde K \Phi\pa _{\infty,q}\le   
\e^{-2}\int _{t^-}^t ds \pa N_{s}^t\tilde K N_{s^-}^s h\pa _{\infty,q} +
\\ &
\e^{-2}\int _{t^-}^t ds\int
_{s^-}^s ds'\Bigg[\pa N_s^t\tilde K N_{s'}^s H'\pa _{\infty,q} +\e^{-2}\pa
N_s^t\tilde K N_{s'}^s \tilde K \Phi\pa_{\infty,q}\Bigg]
}
\Eq(h.2)$$
In this way all the terms we got in \equ(h.2) are analogous to terms already
discussed in the first part of the proof but the term containing $h$. The
problem with this term is that it can be estimated in terms of the $L_\infty$
norm of $\Phi$ but we need to gain a small factor and $N_{s^-}^s$ cannot
provide it.  Hence we have to repeat the previous argument  and use \equ(4.31)
to represent $\Phi$ on the boundary in terms of the function evaluated on the
point on the boundary reached after a finite amount of time. 

We get
$$\eqalign{&
\e^{-2}\int _{t^-}^t ds \pa N^t_{s}\tilde K N^s_{s^-} h\pa
_{\infty,q}\le 
\e^2\pa \zeta\pa _{q} +\\&
\e^{-2}\int _{t^-}^t ds \pa N^t_{s}\tilde K N^s_{s^-}\big[
\tilde M^{-1/2}[{\overline M}^+\a^+_R+\a^-_R{\overline M}^-]\big]\pa
_{\infty,q}+\\&\e^{-2}\int _{t^-}^t ds \int _{s^-}^s d\tau\Big[
\pa N^t_{s}\tilde K N^s_{s^-} N_{s_-}^\tau H'\pa
_{\infty,q}+\e^{-2}\pa N^t_{s}\tilde K
N^s_{s^-}N_{s_-}^\tau\tilde K\Phi\pa _{\infty,q}\Big]
}\Eq(h.4)$$ By using the properties of $N_s$ and $\tilde K$ and
the bounds on the integral  in the exponential as discussed
before the second term in
\equ(h.4) is bounded by
$\e^2\pa\Phi\Pa_{\infty,q}$. 

Finally we get
$$\eqalign{
\sup_{(x,y)\in \Bbb T^2}&|\a^\pm_R|\le C\e^2\pa\Phi\Pa_{\infty,q}+\e^2\pa
\zeta\pa _q+ \int _{t^-}^t ds\Pa N_s H'
\Pa_{\infty,q}+\\&
\e^{-2}\int _{t^-}^t ds\int_{s^-}^s ds'\Bigg[\pa N_s^t\tilde KN_{s'}^s
H'\pa_{\infty,q} +\e^{-2}\pa N_s^t\tilde K N_{s'}^s \tilde K
\Phi\pa_{\infty,q}\Bigg]
}
\Eq(h.5)
$$
Now we can apply all the previous argument to get the following estimate for 
the boundary term in \equ(4.41)
$$\eqalign{
\Pa N_{t^-}^t &h\pa_{\infty,q}\le \sup_{(x,y)\in \Bbb T^2}[|\a^+_R| +|\a_R^-|]
\le\\& C\e^2\pa \zeta\pa _{q}+C\e^2
\sup_t\pa   H'\pa _{\infty,q}\
+C\e^{-d/2}\sup_t \pa 
\Phi\pa _{2,2}}
\Eq(h.6)
$$

Finally by \equ(4.25.1) we have
$$\eqalign{
\sup_{0\le t\le \bar t}\pa \Phi\pa_{\infty,q}&\le C\e^2 \sup_{0\le t\le \bar
t}\pa \zeta\pa _{q}+C\e^2\sup_{0\le t\le \bar t}\pa  H'\pa_{\infty,q}\\&+
C\e^{-d/2}\sup_{0\le t\le \bar t} \pa \tilde D \pa
_{2,2}+\e^{-d/2}\sup_{0\le t\le \bar t}\pa\zeta\pa_{2}
}
\Eq(4.42)$$
This concludes the proof of Theorem 4.4.

\vskip.5cm

To get the  $L^\infty$ bound for $\Phi$ we need to estimate the $\pa \cdot\pa _{\infty,\infty}$
norm in terms of the $\pa \cdot
\pa _{2,2}$ of $\Phi$.  The use Lemma 4.5 allows to prove the following

\vskip.2cm

    \noindent{\bf Theorem 4.6}
    \vskip.2cm
    \noindent{\it Define
    $$|f|_r=\sup_{x\in\Lambda}\sup_{v\in \Bbb R^3}|(1+|v|)^r f|.$$
    Let $\Phi$ be the solution of \equ(4.21), \equ(4.21.1). 
    Then if $r>3$ 
    $$ \sup_{0\le t\le \bar t}|\Phi|_r\le
    C\e^{-d/2}\Bigg(\sup_{0\le t\le \bar
    t}|\tilde D|_r + \sup_{0\le t\le \bar t} |\zeta|_r\Bigg)$$ }
    \vskip.2cm
\vfill\eject
{\it Proof}

By \equ(4.31) we have
$$|\Phi|_0\le |N_{t^-} h|_0 +\int_{t^-}^t ds |N_s H'|_0 +\e^{-2}\int_{t^-}^t ds |N_s \tilde K
\Phi|_0$$
By \equ(4.32.3) and Lemma 4.5 (used with $r=+\infty$ and $s$ 
replaced by $q$), for $q>{d\over 2}$ we have 
$$|\Phi|_0\le C\e^2\sup_{0\le t\le\bar t}| h|_0 +C \sup_{0\le t\le \bar t}| H'|_0
+C\sup_{0\le t\le \bar t} |\Phi|_{\infty,q}
$$
By \equ(4.42) we get
$$
\eqalign{
|\Phi|_0\le C\sup_{0\le t\le \bar t}| h|_0 &+C\e^2 \Bigg(\sup_{0\le t\le
\bar t}| H'|_0 +\sup_{0\le t\le \bar t}\pa H'\pa_{\infty,q}+C\e^2\sup_{0\le
t\le\bar t} \pa\zeta\pa _{\infty,q}\Bigg)\\& +C\e^{-d/2}\Bigg(\sup_{0\le t\le
\bar t}\pa\tilde D\pa_{2,2}+\sup_{0\le t\le
\bar t} \pa \zeta\pa_{2}\Bigg)
}
$$
Because of the factor $(1+|v|)^r$ in the definition of $|\cdot|_r$ and of
the boundedness of the space domain, if $r>3$ we have 
$$\pa  f\pa _{\infty,q}\le C|f|_r,\qquad \pa f\pa _{2,2}\le C |f|_r.$$
Hence
$$\sup_{0\le t\le \bar t}|\Phi|_0\le C\sup_{0\le t\le \bar t} |
h|_0+ C
\e^{-d/2}\Bigg(\sup_{0\le t\le \bar t}| \tilde D|_{r}+
\sup_{0\le t\le \bar t}|\zeta|_{r}\Bigg)+C\e^2\sup_{0\le t\le \bar
t}|H'|_r.\Eq(4.44.1)$$

The boundary term $|h|_0\le C|\a^+_R|_{\infty} +|\a^-_R|_{\infty}
+|\zeta|_{\infty}$ has been  estimated before by using \equ(h.6). The term
containing $H'$ will be bounded using the Grad estimate
$$|M^{-1/2}Q^+(f,g)|_{r}\le C|M^{-1/2}f|_{r}|M^{-1/2}g|_{r},
\Eq(4.43)$$
where $Q^+$ is the gain term of the collision operator.

By using \equ(4.43)  and the definitions of $\tilde K^i, i=1,2$ we get
$$\pa \tilde K^1\Phi\pa _{r}\le  C\pa M^{-1/2}f_1\pa _{r}\pa \Phi\pa _{r}$$
$$\pa \tilde K^2\Phi\pa _{r}\le  C\pa M^{-1/2}\sum_{n=2}^7f_n\pa_r\pa \Phi\pa
_r.$$ 
Finally the regularity property of the hydrodynamic solution and Theorem 4.1
give
$$\pa \tilde K^1\Phi\pa _{r}\le C\pa \Phi\pa_r,\qquad \pa \tilde
K^2\Phi\pa _r\le C\pa \Phi\pa_r\Eq(4.37)$$

Hence,  for $\e$ small

$$\sup_{0\le t\le \bar t}|\Phi|_0\le C \e^{-d/2}\Bigg(\sup_{0\le t\le \bar t}|
\tilde D|_{r}+\sup_{0\le t\le \bar t}|\zeta|_r\Bigg).\Eq(4.44)$$

Finally we  improve the $|\cdot|_0$ norm to $|\cdot|_r$ norm by means of the 
Grad estimate and the representation \equ(4.31) of $\Phi$
$$|\Phi|_r\le C\left(\e^2 \sup_{0\le t\le \bar t}|H'|_r +\sup_{0\le t\le \bar
t}|\zeta|_{r}+ \sup_{0\le t\le \bar t} |\Phi|_{r-1}\right)$$ 
By \equ(4.43), iterating the previous inequality we get,
for $\e$ small, 
$$\sup_t|\Phi|_r\le C\e^2 \sup_{0\le t\le \bar t}|\tilde D|_r +
C\sup_{0\le t\le \bar t} |\Phi|_{0}+C \sup_{0\le t\le \bar t}| 
\zeta|_{r}.\Eq(4.45)$$

By putting together \equ(4.44) and \equ(4.45)  we eventually get
$$\sup_{0\le t\le \bar t}|\Phi|_r\le C\e^{-d/2}\Bigg(\sup_{0\le t\le \bar t}|
\tilde D|_{r}+C\sup_{0\le t\le \bar t}|\zeta|_{r}\Bigg).$$ 
so proving the Theorem.

\vskip.3cm

\noindent{\it Non linear case}

We conclude our discussion by proving the following
\vskip.2cm
\noindent{\bf Theorem  4.7}
\vskip.2cm
\noindent{\it There is $\e_0$ such that, if $\e<\e_0$, for $0<t\le
\bar t$ there is a unique solution to  the initial boundary value problem
\equ(3.14), \equ(3.19), \equ(3.20) verifying the following: for any positive
integer $\ell$ there is a constant
$c>0$ such that 
$$|R|_{\ell,h} \le C \e\Eq(2.20.66)$$
for any $h\le 1/(4 T_-)$.} 
\vskip.2cm
\noindent{\it Proof  }
\vskip.1cm

The equation we have to solve is

$$\pt_t \Phi+ \e^{-1}v\cdot\nabla \Phi +\u G\cdot\nabla_v \Phi=
\e^{-2} {\tilde L} \Phi +\e^{-1}{\tilde L}^1 \Phi +{\tilde L}^2 \Phi+ \tilde D$$
with 
$$\tilde D=\e^2 M^{-1/2}Q(\sqrt M \Phi,\sqrt M \Phi)+\e^2 M^{-1/2}A$$
and $A$ given by \equ(3.16).

By Theorem 4.6, for $d=3$ we have that
$$\eqalign{
\sup_{0\le t\le \bar t}|\Phi|_r&\le \e^{1/2} C\left(\sup_{0\le t\le \bar t}
|\tilde M^{-1/2}Q(\sqrt {\tilde M}
\Phi,\sqrt {\tilde M}\Phi)|_r+
\sup_{0\le t\le \bar t}|\tilde M^{-1/2}A|_r\right)\\& +\e^{-3/2} C
\sup_{0\le t\le \bar t}|\zeta|_{r}}$$

By \equ(3.49) 
$$\sup_{0\le t\le \bar t}|\Phi|_r\le C\e^{1/2}\left((\sup_{0\le t\le \bar t}
|\Phi|_r)^2 + \e\sup_{1/2}|\tilde M^{-1/2}A|_r\right)+C\e^{-3/2} 
\sup_{1/2}| \zeta|_{r}\Eq(4.46)$$

By the arguments  in [\rcite{ELM1}] we then get 
$$\sup_t|\Phi|_r\le  \e^{1/2} C\sup_{0\le t\le \bar t} |\tilde M^{-1/2}A|_r+C
\e^{-3/2}\sup_{0\le t\le \bar t}|\zeta|_{r}$$ 
The term $\zeta$ decays exponentially fast in $\e$. 
Hence, by  Theorem 4.1 the estimate \equ(2.20.66) follows.

\vskip.7cm
\numsec=5
\numfor=1
{\bf 5. The stationary case}
\vskip.5cm
The main difference between the results for the time dependent case of previous
Section and those for the stationary case we are going to present is in the
restriction to small values of the Rayleigh number, we need to deal with the
stationary problem. Therefore, at the hydrodynamical level we are confined to the
purely conductive solution. We hope to be able to extend our method to 
the convective solutions which appear for larger values of the Rayleigh number.
The proof follows by argument quite similar to those presented in [\rcite{ELM1}],
[\rcite{ELM2}] to which we refer the reader for more details. 
In this section we only give a sketch of the proof.

We start by recalling the stationary setup. We look for one-dimensional
solutions, namely for solutions not depending on $x$ and $y$ so that the
equation we consider is 

$$v_z\pt_z f^\e + \e \u G\cdot\nabla_v f^\e= {\e}^{-1} Q(f^\e,f^\e)\Eq(5.1)$$  

The boundary conditions are:
$$ f^\e(-1,v )=\alpha_- \overline M_-(v), \phantom{....}v_z>0,\Eq(5.2)$$
$$
f^e(1,v)=\alpha_+ \overline M_+(v), \phantom{....}v_z<0,\Eq(5.3) $$
with  
${\overline M_\pm}(v)$ given by \equ(3.5) and $\alpha_\pm$ now independent on $x$, 
$y$ and $t$, given by \equ(3.7), so that $f^\e$ satisfies \equ(3.6).

We construct the solution in the form \equ(3.8) with $f_n$ to be determined according
to a  bulk and boundary layer expansion.  These terms are
computed as in the time-dependent case and a theorem similar to
Theorem 4.1 can be stated also in this case. We only discuss the
remainder equation because its solution requires a different 
technique. This equation has the form
$$v_z{\pt\over \pt z} R -\e G{\pt\over \pt v_z}
R= {1\over \e}{{\cal L}} R + {{\cal L}}^{(1)} R + \e
{\cal L}^{(2)} R +
\e^3 Q(R,R) + \e^3 A \Eq(5.8)$$
with ${\cal L}^{(1)} $ and ${\cal L}^{(2)} R$  defined in \equ(3.15). 
Moreover, $A$ is given by 
$$\eqalign{ A =-&   v_z{\pt\over\pt z} B_7+G{\pt\over\pt v_z} (B_6 +\e B_7) +
(G^0+G^-){\pt\over\pt v_z} [(b^+_6 +\e b^+_7)]+
\\&(G^0+G^+){\pt\over\pt v_z}[(b^-_6+\e b^-_7)]+\sum_{\st k,m\ge 1 \atop\st k+m\ge 8 }
\e^{k+m-8} Q(f_k,f_m). 
}
\Eq(5.9)$$
The boundary conditions on $R$ are given by \equ(3.19), \equ(3.20). $R$ satisfies the
normalization condition \equ(3.20.1) and the vanishing flux condition
$$\int dv v_z R(z,v)=0 \qquad\hbox{for} \ z\in [-1,1].\Eq(5.13)$$

The theorem below summarizes the results about the existence of
stationary solutions.
\vskip.2cm
\noindent{\bf Theorem 5.1}
\vskip.2cm
\noindent{\it Let $M$ be the Maxwellian with
parameters $\bar\rho$, $T_-$ and vanishing mean velocity. Put
$$f_1= 
M\Big({\hat r\over \bar\rho}  + {v^2- 3 T_-\over 2 T_-^2}\hat\theta \Big)\Eq(5.15)$$
with $\hat r$ and $\hat \theta$ the thermal conduction solution of the
OBE corresponding to the temperatures $T_-$ and $T_+=T_-(1-2\e\l)$, namely 
$$\hat\theta=T_-\lambda(1+z),\qquad \hat r=-\bar\rho[{G\over T_-}-\lambda]z.$$

Then there are $\lambda_0>0$ and $\e_0>0$ such that, if $\lambda<\lambda_0$ and
$\e<\e_0$, there is a stationary solution to the boundary value
problem above such that
$$|f^\e-(M+\e f_1)|_r\le C\e^2  \Eq(5.16)$$
}

\noindent{\it Sketch of the proof.}

We follow the strategy of the previous section: first we get an $L_2$ bound and then 
the $L_\infty$ bound. In the present case we cannot use the initial condition to
satisfy the normalization condition. Therefore, we satisfy the conditions
\equ(3.20.1)  and \equ(5.13) by choosing the constants $\a^\pm_R$ along the lines of
[\rcite{ELM2}].

Observe that \equ(5.13) is satisfied for any $z\in[-1,1]$, once it is satisfied at
one point, because $\int dv v_z R(z,v)$ does not depend on  $z$ in consequence of
\equ(5.8). 

We write $R$ as 
$$R = I(\Phi) \tilde M + \sqrt {\tilde M}\Phi \Eq(5.17)$$
with 
$$ I(\Phi)= -\int_{-1}^1 dz \int_{\Bbb R^3} dv R(z,v),
\Eq(5.18)$$
so that \equ(3.20.1)  is satisfied.  Therefore we have
$\alpha^-_R=(T_-/2\pi)^{-1/2}\bar\rho^{-1} I(\Phi)$.  It is easy to check (see
[\rcite{ELM2}]) that the function $\Phi$ has to solve the following boundary value
problem: 
$$\eqalign{&
v_z {\pt \Phi\over \pt z } -\e G {\pt \Phi\over \pt v_z}
={1\over \e}{\tilde  L} \Phi + {\tilde L}^1 \Phi
+ {\Cal N} \Phi + \e^2 \ti Q(\Phi,\Phi) + \e^2  A,
\\&\Phi(-1,v)=\tilde M^{-1/2} \zeta^-  \phantom{....} v_z>0,\\&
\Phi(1,v)=\beta_R \overline M_+(v) \tilde M^{-1/2}+\zeta^+\tilde
M^{-1/2}\phantom{....} v_z<0\\
&\int d\/v v_z \Phi\tilde M^{1/2}=0
}\Eq(5.19)$$ 
where $\zeta^\pm =-\sum_{ n=1} ^6  \e ^{n-3}\ga^\pm _{n,\e}$
and $\beta_R=\a^+_R-\a^-_R$. The linear operator ${\Cal N} \Phi$ is defined by 
$${\Cal N} \Phi= \e{  L}^2 \Phi + I(\sqrt {\tilde M}\Phi) \Big[\sum_{n=2}^6 \e^{n
}{\tilde L} f_n \Big].\Eq(5.21)$$
The  non linear term is given by
$$\ti{Q}(\Phi,\Phi) = {1\over\sqrt M}{Q}(\sqrt M\Phi,\sqrt M\Phi) + 
2 I(\sqrt {\tilde M}\Phi) {\tilde L}\Phi.
\Eq(5.21.1)$$

In this way there is no normalization condition on the function 
$\Phi$. The quantity
$\a^-_R$ represents both the outgoing flux of $f_R$ at $z=-1$ and the integral of $R$
over $z$ and $v$. The constant $\beta_R$ is determined so that 
$R$ satisfies condition \equ(5.13) at the point $z=1$, i.e.
$$\beta_R =\int_{v_z>0}dv v_z R(1,v)+\int_{v_z<0}dv
v_z\zeta^+.\Eq(5.22)$$

To construct the solution of \equ(5.19), we first consider the 
following linear boundary value problem: given $D$ on
$[-1,1]\times \Bbb R^3$ and $\zeta^\pm$ on $\{v\in \Bbb R^3 s.t.\, v_z\lessgtr 0\}$, 
find $R$  satisfying
$$v_z {\pt \Phi\over \pt z }-\e G {\pt \Phi\over \pt v_z}={1\over \e}{\tilde
L}\Phi+{\tilde L}^1\Phi + {\Cal N} \Phi +  \tilde D, \Eq(5.23)$$
and the last three conditions \equ(5.19).
 
As usual we introduce $\hat \Phi$  and $\bar\Phi$ the hydrodynamic and the
non-hydrodynamic part of $\Phi$ respectively. Multiplying \equ(5.23) by $\Phi$  and
integrating on velocities, we have
$${1\over 2}\langle v_z\Phi^2\rangle= {1\over \e}\langle\bar\Phi\tilde L
\bar\Phi\rangle+\langle\bar\Phi\tilde L^1 \Phi\rangle+ 
\langle\bar\Phi{\Cal N}
\Phi\rangle+\langle \tilde D \Phi\rangle$$
By integrating over $z$, using \equ(LB), \equ(BL1) and \equ(BL2) we get
$${1\over 2}\langle v_z\Phi^2\rangle(1)-{1\over 2}\langle v_z\Phi^2\rangle(-1)\le -
C{1\over
\e}\pa \bar\Phi\pa ^2 + C[\lambda +\e]\pa\bar \Phi\pa \ \big[\pa \hat\Phi\pa  +\pa 
\bar\Phi\pa \big] + \pa \tilde D\pa \ \pa \Phi\pa.$$ 
Here $\pa\Phi\pa^2=\int_{-1}^1d\/z\int d\/v \phi^2\nu$.
One can check (see [\rcite{ELM2}]) that the l.h.s. of this
inequality is positive. Therefore we get
$$\pa \bar\Phi\pa ^2 \le C \e \lambda \pa \bar \Phi\pa \ \pa \hat\Phi\pa  +\e\pa
\tilde D\pa \ \pa \Phi\pa $$ 

Using the inequality $xy\le kx^2 +y^2/4k$, with $x=\pa\bar\Phi\pa$,
$y=\e\pa\hat\Phi\pa$ and a suitably small $k$, we find
$$\pa \bar\Phi\pa ^2 \le C \e^2 \lambda  \pa \hat\Phi\pa ^2 +\e\pa \tilde D\pa \ \pa
\hat
\Phi\pa +C\e^2\pa\tilde D\pa^2\Eq(5.24)$$

To bound the hydrodynamical part, we multiply \equ(5.23) by $v_z \Psi_i$,
$\Psi_i=\sqrt{\tilde M}\chi_i$, $i\ne 2$  and integrate over $[-1,z]\times\Bbb R ^3$.

Denoting  $p_i(z)=<v_z \Psi_i \Phi>$, we get
$$p_i(z)= p_i(-1) +\int_{-1}^z dz'\int dv v_z\Psi_i\Big[{1\over \e}\tilde L \bar \Phi
+ \tilde L^1
\Phi +\e{\Cal N}\Phi +\e G{\pt\over\pt v_z}\Phi\Big]$$

Following [\rcite{ELM1}], pag.68--69, we can prove that $p_i$ have the following
estimate:
$$|p_i(-1)|\le \big[C \lambda \pa \bar \Phi\pa \ \pa \hat\Phi\pa  +{1\over\e}\pa
\tilde D\pa \ \pa \Phi\pa \big]^{1\over 2}.$$
Using \equ(5.24) this inequality becomes
$$p_i(z)=  \big[C  \lambda \pa \bar \Phi\pa \ \pa \hat\Phi\pa  +{1\over\e}\pa
\tilde D\pa\ 
\pa \Phi\pa \big]^{1\over 2} +C\Big[{1\over \e}\pa  \bar \Phi\pa  +(\lambda +\e)\pa
\hat
\Phi\pa  \Big]\Eq(5.25)$$

Since $\langle v_z\Phi\sqrt{\tilde M}\rangle=0$, we can decompose $\hat \Phi$ as
$$\Phi=\sum_{i\ne 2}h_i\Psi_i.$$ 
Therefore
$$p_i(z)=\sum_{i\ne 2}h_jB_{ij}+ <v_z^2\Psi_i\bar\Phi>$$
with $B$ a non-singular matrix. This allows to estimate $h_i$ and hence   $\hat \Phi$ as
$$ \pa\hat \Phi\pa\le C\Big[{1\over\e}\pa \bar \Phi\pa +{1\over\e}\pa D\pa  +\lambda
\pa\hat\Phi\pa \Big]\Eq(pivot)$$
so that for $\lambda$ small enough  we get 
$$ \pa \hat \Phi\pa \le C\Big[{1\over\e}\pa \bar \Phi\pa +{1\over\e}\pa D\pa 
\Big]\Eq(5.26)$$ 
Combining \equ(5.24) and \equ(5.26) we get
$$\pa \bar \Phi\pa \le C\pa D\pa ,\qquad \pa \hat \Phi\pa \le C{1\over\e}\pa D\pa 
\Eq(5.27)$$
Note that the only point where we need $\l$ small is to pass from \equ(pivot) to
\equ(5.26).
\vskip.3cm
\noindent{\it $L^\infty$ bounds.}
\vskip.2cm
To write the integral form of the linear equation \equ(5.23) we follow the approach
in [\rcite{CEM}]. The notation is as follows:

\noindent The ``total energy'' is $E(z,v)= {v_z^2/2}+V(z)$ with $V(z)=\e G(z+1)$. The
lines with fixed $E$ are the characteristic curves of the equation
$$v_z {\pt f \over \pt z}-\e G{\pt f \over \pt v_z}=0$$
For $E(z,v)>V(z')$ we define
$$v_3(v,z,z')=\sqrt{2E(z,v) -2V(z')}$$
$$v(z,z')=(v_x,v_y,v_3(v,z,z')) $$
$${\cal R}v=(v_x,v_y,-v_z);\quad  {\cal R}v(z,z')=(v_x,v_y,- v_3(v,x,x'))$$
Moreover call  $z^+(z,v)$ the function implicitly defined by the equation 
$$\quad E(z,v)=V(z^+)$$
Finally put
$$\Omega_{z,z'}(v)=\int_{z'}^z d\eta{\nu(\eta,v(z,\eta))\over
v_3(v,z,\eta)};\quad  {\cal R}\Omega_{z,z'}(v)=\int_{z'}^z
d\eta{\nu(\eta,Rv(z,\eta))\over v_3(v,z,\eta)}$$
\vskip.3cm
\noindent Consider the equation
$$v_z {\pt \Phi\over \pt z }-\e G {\pt \Phi\over \pt v_z}+{1\over \e}{\tilde
\nu}\Phi={1\over
\e}H \Eq(5.27.1)$$
with  boundary conditions
$$f(\pm 1,v)=h(v)^\pm, \quad v_z\gtrless 0$$
The solution of \equ(5.27.1)  can be written in an
integral form as:
\vskip.1cm
\noindent for $v_z >0 $:

$$\eqalign{
f(z,v)&=h^-(v(z,-1))\exp -{1\over \e}\Omega_{z,-1}(v) \\
&+\int_{-1}^zdz' {1\over \e}{1\over
v_3(v,z,z')}H(z',v(z,z'))\exp -{1\over \e}\Omega_{z,z'}(v),
}
\Eq(5.28)$$
\noindent for $ v_z<0 \and E<V(1) $:
$$\eqalign{
f(z,v)&= h^-(v(z,-1))\exp -[{1\over \e}\Omega_{z^+,-1}+{1\over
\e}{\cal R}\Omega_{z^+,z}]\\ &+\Big[\int_0^{z^+}dz' {1\over
v_3(v,z,z')}H(z',(v(z,z')))\exp -[{1\over \e}\Omega_{z^+,z'}+{1\over \e}{\cal
R}\Omega_{z^+,z}]\\ &+\int_z^{z^+} dz' {1\over \e}{1\over
v_3(v,z,z')}H(z',Rv(z,z'))\exp -{1\over \e}{\cal R}\Omega_{z,z'}\Big ],
}
\Eq(5.29)$$
\noindent for $ v_z<0 \and E>V(1)$:
$$\eqalign{
f(z,v)&=\exp -{1\over \e}\Omega_{1,z} (v) h^+(v(z,1))\\
&+\int_z^1 dz' {1\over \e}{1\over
v_3(v,z,z')}H(z',v(z,z'))\exp-{1\over \e}\Omega_{z',z}. 
}\Eq(5.30)$$
\vskip.3cm
We can write the previous formulas in a compact form as
$$f= Vh +TH\Eq(5.31)$$
where 
$$
\eqalign{
&V_1f(z,v)=\chi(v_z>0) h^{-}(v(z,-1))\exp
-{1\over\e}\Omega_{z,-1}(v),
\\&V_2f(z,v)=\chi(v_z<0)\chi(E<V(1)) h^{-}(v(z,-1))\exp
-{1\over\e}[\Omega_{z^+,-1}+{\cal
R}\Omega_{z^+,z}],\\&
V_3f(z,v)=\chi(v_z<0)\chi(E>V(1)) h^{+}(v(z,1))\exp
-{1\over\e}\Omega_{1,z}\\&
Vf=\sum_{i=1}^3V_if.}$$
The definition of $T$ is given in a similar way.

By slightly modifying  the proof in [\rcite{CEM}] to take into account the factor 
$\e$ it is possible to prove the following Lemmas (see also [\rcite{ELM1}]).

\goodbreak
\noindent{\bf Lemma 5.2} 
\vskip.2cm 
 \noindent{\it For any integer $r\ge 0$ there is a constant $c$ such that the
integral operator $T$ satisfies the following inequality,
$$|T\  H|_r \le c \Big | {H \over \nu}\Big | _r .
\Eq(5.32)$$
}
\vskip.1cm
Here
$$|f|_r=\sup_{z,v}(1+|v|)^r |f(z,v)|.$$
\vskip.2cm
\goodbreak
\noindent{\bf Lemma  5.3} 
\vskip.2cm 
\noindent{\it For any $\delta >0$ and for any $r\ge2$  there is  a constant $C_\de$ 
such that
$$N(T\ H) \le  {1\over\sqrt\e} C_{\de} \pa \nu ^{-1/2}H \pa  + \de |H|_r. \Eq(5.33)$$
}
Here
$$N(f) = \sup_{z\in [-1,1]} \left( \int_{\Bbb R^3} dv |f(z,v)|^2 \ 
\right)^{1/2}. $$

We write \equ(5.22) in the form \equ(5.27.1) with
$$H=\tilde K\Phi +\e {\tilde  L} \Phi  +\e {\tilde L}^1 \Phi+\e  {\Cal N} \Phi +\e
D\Eq(a)$$ and $h^\pm=\Phi(\pm 1,v)$ given by \equ(5.19). 
Combining these Lemmas and using the properties of the
operator
$L$ one gets
\vskip.4cm
\goodbreak
\noindent{\bf Theorem 5.4}
\vskip.2cm
\noindent{\it 
There exists a constant $C$ such that for any $r\ge 3$ the solution of
\equ(5.23) verifies
$$|\Phi|_r\le C{1\over
\sqrt\e}\pa \nu^{-1/2}\Phi\pa _2)+{1\over\sqrt\e}\pa \e\nu^{-1/2}\tilde D\pa
+\e|\tilde D|_r+|Vh|_r\Eq(5.34)$$ }
Noting that $\pa \nu^{-1/2}D\pa \le |\nu^{-1}\tilde D|_3$ and using \equ(5.26) we get
$$|\Phi|_r\le C{1\over
\sqrt\e}|D|_{r-1}+|Ah|_r\Eq(5.35)$$ 
Finally \equ(5.35) implies
$$|R \tilde M^{-1/2}|_r\le C\sqrt\e |\tilde M^{-1/2}D|_{r-1}+|Vh|_r\Eq(5.36)$$

The term containing $h$ involves $\b_R$ which still depends on $\Phi$.
To estimate it, we follow the method in [\rcite{ELM2}]. 
 
Equation \equ(5.31) allows to express $\beta_R$ in terms of
$T H$ and the restriction of $\Phi(-1,v)$ to $v_z >0$. We have
the estimate
 $$\eqalign{ |\int_{v_z>0} v_z M_*^{1/2} h(1,v)| &= |\int_{v_z>0}v_z 
M_*^{1/2}\/\{
h^-(v(z,-1))\exp -{1\over \e}\Omega_{z,-1}(v) \\
&+\int_{-1}^zdz' {1\over \e}{1\over
v_3(v,z,z')}H(z',v(z,z'))\exp -{1\over \e}\Omega_{z,z'}(v),
\}|\\
&\le |s| + |\int_{v_z>0}v_z M^{1/2}\/ TH|,
}\Eq(5.37)$$
 with $s_\pm=\zeta^\pm M ^{-1/2}$ and $|s_\pm|=\sup_{v_y\lessgtr
0,z}|s_\pm(v)|$. 
By  \equ(5.31), using the Schwartz inequality and
$$\int_{-1} ^1 dz {1\over \e}{1\over
v_3(v,1,z)}H(z,v(1,z))\exp -{1\over \e}\Omega_{1,z}(v)<C\int_0^\infty dz\exp[-z]<C,$$
we get
$$\eqalign{
|\int_{v_z>0}v_z M^{1/2} T H\big |&\le 
\e ^{-1/2} C\int_{v_z>0} v_z M^{1/2} \Big [\int_{-1}^1 dy
\nu^{-1}{1\over
v_3(v,1,z,)} H^2\Big ]^{1/2}\\
&\le \e ^{-1/2} C\int_{v_z>0} |v_z|^{1/2} M^{1/2} \Big [\int_{-1}^1 dy
\nu^{-1} H^2\Big ]^{1/2}
}$$ 
Finally, using again the  Schwartz inequality and recalling the expression of
$\beta_R$ in \equ(5.22) we get
 
$$\beta_R \le c [\e ^{-1/2} \pa  \nu^{-1}H\pa _2 +|s|].\Eq(5.38)$$
Using now the form  of $H$ in \equ(a)
$$\beta_R \le c [\e ^{-1/2} \pa  \nu^{-1}\Phi\pa _2 +\pa  \e D\pa _2+|s|].$$

As a consequence \equ(5.36) becomes
$$|R \tilde M^{-1/2}|_r\le C\sqrt\e
|\tilde M^{-1/2}D|_{r-1}+C|\tilde M^{-1/2}\zeta^+|_r+|\tilde
M^{-1/2}\zeta^-|_r\Eq(5.39)$$ which concludes the analysis of
the linear case. The non-linear problem is dealt with by a fixed
point argument and the final result is

\vskip.4cm
\goodbreak
{\bf Theorem 5.5}

{\it 
There exist $\lambda>0$, $\e_0>0$  and a constant $C$ such that, 
if $\lambda<\lambda_0$
and
$\e<\e_0$, there is a  solution to the boundary value
problem \equ(5.8), \equ(3.19), \equ(3.20), \equ(5.13)  such that for any $r\ge 0$ 

$$|M^{-1/2}R|_r\le C\e^{3/2}|M^{-1/2}A|_r\Eq(5.40)$$
}

This concludes the proof of Theorem 5.1.
\vskip.6cm

{\bf Acknowledgements}
\vskip.2cm
Research supported in part by AFOSR Grant 95-0159, and by 
CNR-GNFM and MURST. We also wish to thank the warm hospitality
of the IHES, Bures-sur-Yvette, where some of this work was done.

\newpage
\numfor=1

{\bf Appendix}

\vskip.3cm

In this appendix we show how to get $L_2$ estimates for the
boundary terms. We will
prove \equ(4.24.a). To this end it is enough to prove 

$$S^+:= \e\int_0^{t_1}dt\int_{\Bbb T^2} dxdy\int_{v_z>0}dv v_z
|\Phi|^2(x,y,1,v;t)\le \e^4
\int_0^{t_1}dt\pa{H_t\over\sqrt\nu} \pa^2 +\sup_{0<t\le
t_1}\zeta(\,\cdot\,,t)|^2 
\Eqa(A.1)$$
where $H_t(\,\cdot\,):=H(\,\cdot\,,t)$ is defined in \equ(4.34). In fact by
substituting the expression of
$H_t$ we get
$$\eqalign{
\e^4 &\pa \nu^{-1/2} H_t\pa^2\le  \pa \nu^{-1/2}\tilde K\Phi_t\pa^2
+\e^2\pa{\nu^{-1/2}\tilde K}^1\Phi_t\pa^2
\\&+\e^4\pa{\nu^{-1/2}\tilde K}^2\Phi_t\pa^2 +\e^4 \pa\tilde D_t\pa^2\le
C \pa \Phi_t\pa^2 +\e^4 \pa\tilde D_t\pa^2
}
$$
so proving \equ(4.24.a).

We use 
\equ(4.31) to express the value of the function $\Phi$ in the
point $z=1$.  We remark that the characteristic 
$\f$ which is on the boundary $z=1$ with velocity $v, v_z>0$ at time $t^- -t$ 
had to start at time $0$ either from some point in the bulk or from the boundary
$z=-1$. In the first case  the boundary term $h$ in \equ(4.31) is zero  otherwise we
get $\Phi(x',y',- 1, v';t)$, $v'_z>0$ where 
$$x'= x+v_x(t^- -t),\quad y'= x+v_y(t^- -t),\quad v'=v-G (t^- -t).$$
Hence
$$\eqalign{
|\Phi|^2(x,y,1,v;t)&\le 2|\Phi|^2\big(t^-,x',y',-1,v')\big)\exp\Big\{-2\int_{t^-}^t 
ds{1 \over\e^2}\nu\big(\f_{s-t}(x,y,1,v)\big)\Big\}\\ & +2\Bigg[\int_{t^-}^t ds
H\big(s,\f_{s-t}(x,y,1,v)\big)
\exp\Big\{-\int_{s}^t
ds'{1 \over
\e^2}\nu\big(\f_{s'-t}(x,y,1,v)\big)\Big\}\Bigg]^2.
}
\Eqa(A.3)
$$
By the boundary condition on $\Phi$ we have
$$\Phi\big(t^-,x',y',-1,v')\big)=\a^-_R(t^-,x',y',-1) 
\overline M_-(v')\tilde M^{-1/2}(v')
+\zeta^-(v'),\quad  v'_z>0$$ so that by Schwartz inequality we get
$$\int_{v'_z>0} v'_z dv'|\Phi|^2\big(t^-,x',y',-1,v')\big)\le C(\a^-_R)^2
(t^-,x',y',-1) +C
\int_{v_z>0}v_z dv|\zeta^-|^2$$
But $\a^-_R$ is expressed again in term of the value of $\Phi$ on the boundary 
$z=-1$ and for negative $z$-component of the velocity, namely
$$(\a^-_R)^2=\Bigg[\int_{v_z<0}|v_z|\sqrt{\tilde M} \Phi(t^-,x',y',-1, v)\Bigg]^2\le
C\int_{v_z<0}|v_z|\, |\Phi|^2(t^-,x',y',-1, v)$$

We observe now that the integral 
$$\int_{t^-}^t ds\nu\big(\f_{s-t}(x,y,-1,v)\big)$$ when evaluated on the 
characteristics going from one boundary to another is bounded from below. In fact we
can use the bound on $\nu$ , $\nu(x,v)\ge c_0(1+|v|)$ to check that this integral for
the trajectory perpendicular to the boundary, namely $v_x=v_y=0, v_z>0$ is greater
than $\e$ times the width of the slab and this is the  worst case  
since for the other trajectories is even greater.  As a
consequence, we can estimate the exponential in the first row of
\equ(A.3) as $e^{-C/ \e}$. In conclusion we get
 
$$ \eqalign{ \e &\int_0^{t_1}dt\int dxdy\int_{v_z<0}dv |v_z| 
|\Phi|^2\big(t^-,x',y',-1,v')\big)\exp\Big\{-2\int_{t^-}^t ds{1 \over
\e^2}\nu\big(\f_{s-t}(x,y,1,v)\big)\Big\}\\
&\le C  e^{-C /\e}\Bigg[ \int_0^{t_1}dt\int dxdy\int_{v_z<0}dv |v_z|\,
|\Phi|^2\big(t,x,y,-1,v)\big)+\sup_{0\le t\le t_1}\pa\zeta^-\pa^2\Bigg].
}\Eqa(A.4)$$
To get the previous expression we have replaced  $x',y'$ and
$t^-$  with
$x,y$  and $t$, since the respective Jacobians are equal to one.

In this way we still do not have any explicit estimate of the boundary term in
\equ(A.3), because it contains the values of $\Phi$ at $z=-1$ for
negative
$v_z$ which is  still unknown. However, we can use the representation
\equ(4.31) again to express it back in terms of the function in $z=1$ so to get a 
set of coupled equations for the boundary terms. Hence we consider
$$S^-:=-\e\int_0^{t_1}dt\int_{\Bbb T^2} dxdy \int_{v_z<0}\/dv v_z|\Phi|
^2(x,y,-1, v;t)
\Eqa(A.5)$$
Equation \equ(A.4) implies
$$S^+\le C e^{-C /\e}[S^- +\sup_{0\le t\le t_1}\pa\zeta^-\pa^2] + \e{\cal
H}^+\Eqa(A.6)$$ where 
$$\eqalign{
{\cal H}^\pm
&=\int_0^{t_1}dt\int_{\Bbb T^2} dxdy\int dv |v_z|\Bigg[\int_{t^-}^t ds
H\big(s,\f_{s-t}(x,y,\pm1,v)\big)\\&
\exp\Big\{-\int_{s}^t
ds'{1 \over
\e^2}\nu\big(\f_{s'-t}(x,y,\pm1,v)\big)\Big\}\Bigg]^2
}$$
As above we use \equ(4.31) to find a bound for $S^-$. 
Define 
$$E={|v|^2\over 2\e^2}+{1\over\e}G(z+1).$$
We have:
\vskip.2cm
\item{} for $v_z<0$, $E>0$
$$\eqalign{\Phi &(x,y,-1,v;t)=\Phi\big(t^-,x',y',1,v')\big)
\exp\Big\{-\int_{t^-}^tds{1\over\e^2}\nu\big(\f_{s-t}(x,y,-1,v)\big)\Big\}\\&
+\int_{t^-}^t ds H\big(s,\f_{s-t}(x,y,-1,v)\big)\exp\Big\{-\int_{s}^t ds'{1 \over
\e^2}\nu\big(\f_{s'-t}(x,y,-1,v)\big)\Big\}.
}
\Eqa(A.7)
$$
\vskip.2cm
\item{} For $v_z<0$, $E\le 0$
$$\eqalign{
\Phi &(x,y,-1,v;t)=
\Phi\big(t^-,x",y",-1,v")\big)
\exp\Big\{-\int_{t^-}^tds{1\over\e^2}\nu\big(\f '_{s-t}(x,y,-1,v)\big)\Big\}\\&
+\int_{t^-}^t ds H\big(s,\f '_{s-t}(x,y,-1,v)\big)
\exp\Big\{-\int_{s}^tds'{1 \over\e^2}\nu\big(\f '_{s'-t}(x,y,-1,v)\big)\Big\}.}
\Eqa(A.7.1)$$
where $(x'',y'',-1,v'')=\f_{t^--t}(x,y,-1,v)$ and $\f '$ is the characteristics 
starting from  $-1$ with $v''_z>0$.

These two formulas correspond to the case in which the characteristics starts from 
$z=1$ with velocity $v'_z<0$ and to the case in which it starts from $z=-1$ with
$v''_z>0$ and then come back to the boundary $z=-1$. This second possibility appears
because of the presence of the force:  the kinetic energy is such that the trajectory
does not reach the opposite boundary but instead goes back after a time depending on
the balance between kinetic energy and potential energy. In that case the total
length $\ell$ for a trajectory with $v_x=0=v_y$ is given by 
$$\ell={1\over 2\e G}[v''_z]^2.$$
The other trajectories give a larger value to this integral.

Hence the square of the second term in the first row of \equ(A.7.1), integrated on 
time, space and velocity, is bounded as
$$\eqalign{
B:=&\int_0^{t_1}dt\int_{\Bbb T^2} dxdy  \int_{v_z<0}|v''_z|\Phi|
^2(x'',y'',-1, v'';t)\exp\{-{C\over\e^2}v^2_z\}\\&
=\int_0^{t_1}dt\int_{\Bbb T^2} dxdy  \int_{v_z>0}|v_z|\Phi|
^2(x'',y'',-1, v''_x,v''_y,v_z;t)\exp\{-{C\over\e^2}v^2_z\},
}
\Eqa(A.8)$$
using that $v''_z=-v_z$.  The function $\Phi$ in the  integral  in \equ(A.8)
is relative to $z=-1$ and  $v_z>0$ so that we can
express it in terms of $\a^-_R$ as follows, for $v_z>0$:
$$\eqalign{
 |\Phi|
^2(x'',y'',-1,& v''_x,v''_y,v_z;t)\le {\a^-_R}^2\bar M_-^2\tilde M^{-1} +(\zeta^-)^2, 
\\&
\le C\Bigg[\int_{v_z<0}dv |v_z|\sqrt {\tilde M} \Phi(x'',y'',-1,
v''_x,v''_y,v_z;t)\Bigg]^2\tilde M +{\zeta^-}^2\\&
\le C\tilde M\int_{v_z<0}dv |v_z| |\Phi|^2(x'',y'',-1, v''_x,v''_y,v_z;t)+(\zeta^-)^2
}
\Eqa(A.9)
$$
Equation \equ(A.9) implies for $B$ the following bound
$$\eqalign{
B\le & C\int_0^{t_1}dt\int dxdy\int_{v_z<0}dv |v_z|
|\Phi|^2(x'',y'',-1, v''_x,v''_y,v_z;t)\\&
\int
dv|v_z|M\exp\{-{C\over\e^2}v^2_z\}+C
\int dv|v_z\pa {\zeta^-}|^2\exp\{-{C\over\e^2}v^2_z\} }\Eqa(A.10)
$$
We have that
$$\int
dv|v_z|\tilde M\exp\{-{C\over\e^2}v^2_z\}\le C\e^2
\Eqa(A.11.1)$$
Finally
$$\eqalign{
B&\le \e^2 C\int_0^{t_1}dt\int dxdy\int_{v_z<0}dv |v_z|
|\Phi|^2(x'',y'',-1, v''_x,v''_y,v_z ;t)+\e^2\sup_{0\le t\le
t_1}\pa\zeta^-\pa^2\\&=C\e S^-+\e^2\sup_t\pa{\zeta^-}\pa^2
}
$$

The contribute of the first term in (A.7) to $S^-$ is 
$$\eqalign{
&\e\int_0^{t_1}dt\int dxdy\int_{v_z<0}dv |v_z|
|\Phi|^2(x',y',1, v';t^-)\exp\Big\{-\int_{t^-}^t
ds{1\over\e^2}\nu\big(\f_{s-t}(x,y,-1,v)\big)\Big\}\le\\&
\e\int_0^{t_1}dt\int dxdy\int_{v_z<0}dv |v_z|\big[\tilde M
|\a^+_R|^2(x',y',1;t^-)+|\zeta^+|^2\big]\\&\times
\exp\Big\{-\int_{t^-}^t ds{\nu\over \e^2}\big(\f_{s-t}(x,y,-1,v)\big)\Big\}\le\\&
Ce^{-C/\e}\Bigg[\int_0^{t_1}dt\int dxdy\int_{v_z>0}dv |v_z|
|\Phi|^2(x',y',1, v';t)+\sup_{0\le t\le t_1}\pa \zeta^+\pa^2\Bigg]\\&
=C e^{-C/\e} [S^+ + \e\sup_{0\le t\le t_1}\pa \zeta^+\pa^2]
}
\Eqa(A.11)
$$
Summarizing we can write
$$S^-\le C\e[ S^- + \exp[-C/\e]S^+ + \e \sup_{0\le t\le t_1}\pa \zeta^-\pa^2]
+\e{\cal H}
\Eqa(A.12)
$$
where ${\cal H}={\cal H}^++{\cal H}^-$. 
This implies, for $\e$ small,
$$S^-\le C\e^{-C/\e}[S^+ + \e \sup_{0\le t\le t_1}\pa\zeta\pa^2]
+\e{\cal H}
\Eqa(A.13)
$$

Putting together (A.6) and (A.13) we get
 $$\eqalign{
 &S^+ \le C\left(\e{\cal H}+ \sup_{0\le t\le t_1}\pa \zeta\pa^2\right),\\&
 S^- \le C\left(\e{\cal H}+ \sup_{0\le t\le t_1}\pa \zeta\pa^2\right).
 }
 \Eqa(A.14)
 $$

To conclude the argument we now provide an estimate for  ${\cal H}$. By the  Schwartz inequality
for the integral on $d\/s$, we have
$$\eqalign{
|{\cal H}^\pm|&\le
\int_0^{t_1}dt\int dxdy\int dv |v_z|\int_{t^-}^t ds 
\big|{H\over \sqrt\nu}\big|^2\big(s,\f_{s-t}(x,y,\pm1,v)\big)
\\&
\int_{t^-}^t ds \nu\big(s,\f_{s-t}(x,y,\pm1,v)\big)\exp\Big\{-\int_{s}^t
ds'{2 \over
\e^2}\nu\big(\f_{s'-t}(x,y,\pm 1,v)\big)\Big\}\le\\&
\le C\e^2\int_0^{t_1}dt\int dxdy\int dv |v_z|\int_{t^-}^t ds
\big|{H\over \sqrt\nu}\big|^2\big(s,\f_{s-t}(x,y,\pm1,v)\big)
}
\Eqa(A.15)$$

Consider the term
$${\cal V}^\pm:=\int_0^{t_1}dt\int_{t^-}^t ds\int dv |v_z|
\big|{H\over \sqrt\nu}\big|^2\big(s,\f_{s-t}(x,y,\pm1,v)\big)$$
and the following change of  variables $(t,v_z)\to (\xi,w)$
$$\xi=\xi(t):=\pm 1 +{1\over \e}v_z(t-s)+{1\over 2\e}G(t-s)^2,\qquad w=v_z +G(t-s)$$
whose Jacobian  is $\e|v_z|^{-1}$. Denote also by $t(\xi)$ the inverse of $\xi(t)$. We
have  
$$
{\cal V}^\pm:= \e \int_{\xi(0)}^{\xi(t_1)}d\xi\int_{t^-(\xi)}^{t(\xi)} ds \int dw
\big|{H\over \sqrt\nu}\big|^2\big(s, x(\xi,w),y(\xi,w),\xi,v_y,v_x,w)\big)
$$
Hence 
$$
|{\cal H}|\le \e^3\int_{0}^{t_1} ds\int_{\Omega} dx\int
dv\big|{H\over \sqrt\nu}\big|^2\big(s,x,v\big).
\Eqa(A.16)
$$
which implies the bound \equ(A.1).

\vskip 2cm
\goodbreak
{\bf References.}

\item{[ \rtag{DR}]} P.G. Drazin and W.H. Reid, 
{\it Hydrodynamic instability}, Cambridge Univ. Press, Cambridge (1981).

\item{[\rtag{EM}]} R.Esposito and R.Marra {\it Incompressible fluids on three
levels: hydrodynamic, kinetic, microscopic},   Mathematical
Analysis of Phenomena in Fluid and Plasma Dynamics, RIMS, Kyoto (1993). 

\item{[\rtag{Mi}]} J.M. Mihaljan {\it A rigorous exposition of the
Boussinesq approximation applicable to a thin layer of fluid},  
Astrophys. J., {\bf 136}, 1126--1133 
(1962).

\item{[\rtag{Jo}]}  D. D. Joseph, {\it Stability of Fluid
Motions}, Springer-Verlag, Berlin (1976).

\item{[\rtag{Ca}]}   R. E. Caflisch, {\it The
fluid dynamic limit of the nonlinear Boltzmann equation}, 
Commun. on Pure and Applied Math., {\bf 33 } 651-666 (1980)

\item{[\rtag{DEL}]}   A. De Masi,
 R. Esposito, J. L. Lebowitz, {\it Incompressible
Navier-Stokes and Euler limits of the   Boltzmann equation}, 
Commun. Pure and Applied Math., {\bf 42}, 1189--1214 (1989). 

\item{[\rtag{ELM1}]} R. Esposito, J.L. Lebowitz and R. Marra {\it Hydrodynamic Limit
of the Stationary Boltzmann Equation in a Slab}, Commun. Math Phys., {\bf 160}, 49--80
(1994).

\item{[\rtag{ELM2}]} R. Esposito, J.L. Lebowitz and R. Marra, {\it The Navier-Stokes
limit of  stationary solutions of the nonlinear Boltzmann equation},  
 J. Stat. Phys., {\bf 78}, 389--412 (1995).
 
\item{[\rtag{UA}]} S. Ukai and K. Asano, {\it Steady solutions of the Boltzmann 
equation for a flow past an obstacle, I. Existence}, Arc. Rat. Mech. Anal., {\bf 84},
249--291, (1983).

\item{[\rtag{BCN1}]} C. Bardos, R. Caflisch and B. Nicolaenko, 
{\it Thermal layer Solutions of the Boltzmann Equation}, Random
Fields: Rigorous Results in Statistical Physics. Koszeg (1984),
J. Fritz, A. Jaffe and D. Szasz editors, Birkhauser, Boston
1985.

\item{[\rtag{CEM}]} C. Cercignani, R. Esposito, R. Marra,
{\it The Milne problem with a force term}, preprint,
  (1996).

\item{[\rtag{Bo}]}  J. Boussinesq, {\it Theorie analytique 
de la chaleur},
Gauthier-Villars, Paris (1903).

\item{[\rtag{CIP}]}{C. Cercignani, R. Illner, M. Pulvirenti,
{\it The Ma\-the\-ma\-ti\-cal Theory of Dilute Gases}, Springer-Verlag,
 New York (1994)}

\item{[\rtag{GPS}]}
 Golse, F.,  Perthame, B.,  Sulem, C. 
{\it On a boundary layer problem for the nonlinear Boltzmann
equation}, Arch. Rat. Mech. Anal., {\bf 104}, 81--96 (1988).

\end